\documentclass [journal]{IEEEtran}
\usepackage{algorithm,amsbsy,amsmath,amssymb,epsfig,bbm,mathrsfs,multirow,amsthm,xcolor}
\usepackage{epstopdf}
\usepackage{bbding}
\usepackage{graphicx,caption,subcaption}
\usepackage{setspace}
\usepackage{algorithmic}
\usepackage{fancyhdr}
\usepackage{subcaption,float}
\usepackage[hidelinks]{hyperref}

\newtheorem{theorem}{\textbf{\textsc{Theorem}}}

\newcommand{\tabincell}[2]{\begin{tabular}{@{}#1@{}}#2\end{tabular}}

\allowdisplaybreaks[4]

\ifCLASSOPTIONcompsoc
\usepackage[nocompress]{cite}
\else
\usepackage{cite}
\fi
\ifCLASSINFOpdf
\else
\fi

\usepackage{setspace}

\begin{document}  
\title{Integrated Sensing and Communication from Learning Perspective: An SDP3 Approach}
\author{Guoliang Li, Shuai Wang, Jie Li, Rui Wang, Fan Liu, \\Xiaohui Peng, Tony Xiao Han, and Chengzhong Xu, \emph{Fellow, IEEE}
\thanks{Part of this paper has been presented at the IEEE International Workshop on Signal Processing Advances in Wireless Communications (SPAWC), Lucca, Italy, Sep. 2021 \cite{SPAWC}. \emph{(Corresponding Author: Shuai Wang and Rui Wang.)}

Guoliang Li, Jie Li, Rui Wang, and Fan Liu are with the Department of Electrical and Electronic Engineering, Southern University of Science and Technology (SUSTech), Shenzhen 518055, China (e-mail: \{ligl2020, lij2019\}@mail.sustech.edu.cn, \{wang.r, liuf6\}@sustech.edu.cn).

Shuai Wang is with the Guangdong-Hong Kong-Macao Joint Laboratory of Human-Machine Intelligence-Synergy Systems, Shenzhen Institute of Advanced Technology, Chinese Academy of Sciences, Shenzhen, 518055, China (e-mail: s.wang@siat.ac.cn).

Xiaohui Peng and Tony Xiao Han are with Huawei technologies, Co. Ltd., Shenzhen, China (e-mail: \{pengxiaohui5, tony.hanxiao\}@huawei.com). 

Chengzhong Xu is with the State Key Laboratory of Internet of Things for Smart City (SKLIOTSC), Department of Computer Science, University of Macau, Macau, China (e-mail: czxu@um.edu.mo). 
}
}
\maketitle

\begin{abstract}
Characterizing the sensing and communication performance tradeoff in integrated sensing and communication (ISAC) systems is challenging 
in the applications of learning-based human motion recognition. This is because of the large experimental datasets and the black-box nature of deep neural networks. 
This paper presents SDP3, a Simulation-Driven Performance Predictor and oPtimizer, which consists of SDP3 data simulator, 
SDP3 performance predictor and SDP3 performance optimizer. 
Specifically, the SDP3 data simulator generates vivid wireless sensing datasets in a virtual environment, 
the SDP3 performance predictor predicts the sensing performance based on the function regression method, and 
the SDP3 performance optimizer investigates the sensing and communication performance tradeoff analytically. 
It is shown that the simulated sensing dataset matches the experimental dataset very well in the motion recognition accuracy. 
By leveraging SDP3, it is found that the achievable region of recognition accuracy and communication throughput consists of a communication saturation zone, a sensing saturation zone, 
and a communication-sensing adversarial zone, of which the desired balanced performance for ISAC systems lies in the third one.                                \end{abstract}


\IEEEpeerreviewmaketitle
\section{Introduction}\label{sec:Int}

\IEEEPARstart{I}{ntegrated} sensing and communication (ISAC) is a promising technology for the next generation cellular system and wireless 
local area network (WLAN). It is expected to considerably improve the spectral, energy and hardware efficiencies of wireless systems \cite{ISAC}. Since wireless resource is shared between sensing and communication functionalities in ISAC systems, it is of significant interest 
to investigate their tradeoff relation. However, the analysis of sensing performance could be challenging. 

Generally, most of the sensing tasks can be classified into three categories, including \emph{detection}, \emph{estimation} and \emph{recognition}. 
The detection tasks aim to determine the state of a target, such as presence/absence. 
The detection probability is a typical metric to quantify the detector performance. In \cite{detection_1}, an integrated communication and passive radar system was considered, where a generalized likelihood ratio test (GLRT) was proposed for target detection and the corresponding detection probability was approximated in a closed-form formula. 
Accordingly, the detection probability was maximized subject to a minimum communication rate constraint. Furthermore, the above method was extended to an integrated multi-static radar and communication system with the aim for power allocation optimization in \cite{detection_2}. 

The estimation tasks acquire the useful parameters, i.e., the distance, velocity and angle, from the sensed targets. 
For instance, a dual functional waveform was utilized to minimize the estimation mean square error (MSE) of the target response matrix while ensuring a worst communication performance in \cite{estimation_1}. 
When the closed-form expression of MSE is not attainable, the Cram\'er-Rao bound (CRB) could be adopted for the sensing performance evaluation, which represents the lower bound of the variance of all the unbiased estimators. For example in \cite{estimation_2}, the CRB was minimized subject to the signal-to-interference-plus-noise ratio (SINR) constraint of communication receivers. 
Furthermore, in \cite{estimation_3, estimation_4}, the estimation rate, quantifying the reduction of the uncertainties for the sensing parameters per second, was proposed, then the tradeoff analysis between the estimation rate and the communication rate was provided. 

\begin{figure*}[!t]
\centering
		\includegraphics[width=170mm]{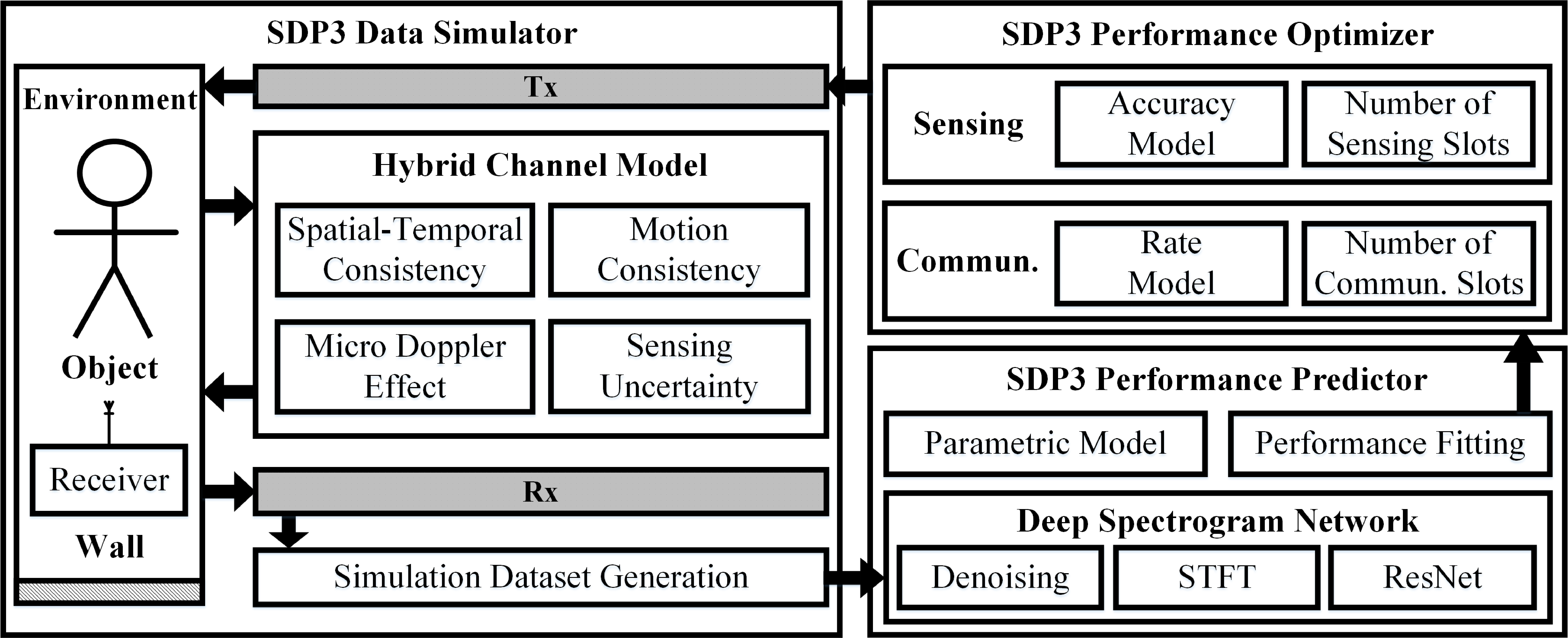}
	\caption{The system diagram of the proposed SDP3 framework.}
	\label{SDP3_diagram}
\end{figure*}

The above two sensing tasks are usually processed in the physical layer, while the recognition tasks are usually accomplished in the application layer aided by the machine intelligence. 
They aim to acquire the semantic information of the sensed targets, e.g., human motion recognition (HMR). 
For example, in \cite{linghao}, the short time Fourier transform (STFT) was adopted to generate spectrograms of the sensing data. 
Then the support vector machine (SVM) was trained for motion classification, whose accuracy was higher than $90\%$.
By replacing the SVM with a convolutional deep neural network (CNN), it was shown in \cite{linghao2} that a higher classification accuracy could be achievable. 
In fact, deep neural network has been adopted extensively in the application of activity recognition \cite{Activity1, Activity2}, hand gesture recognition \cite{Gesture1,Gesture2}, and gait recognition \cite{Gait1, Gait2} with radio wave.
However, most of the HMR tasks are based on deep neural network, where the relation between sensing accuracy and sensing resource (e.g., sensing time and transmission power) can hardly be represented analytically. 
Moreover, the above works rely on the motion datasets generated from time-consuming and labor-intensive experiments. 
As a result, there are the following challenges in the sensing and communication tradeoff analysis with the recognition tasks: 
1) there is no analytical model of recognition accuracy, which maps the sensing resource (e.g., sensing time, power and etc.) to the recognition accuracy; 
2) even for numerical performance evaluation, there is no low-cost method to generate the motion sensing datasets for the training of recognition models.

\begin{figure*}[!t]
\centering
		\includegraphics[width=170mm]{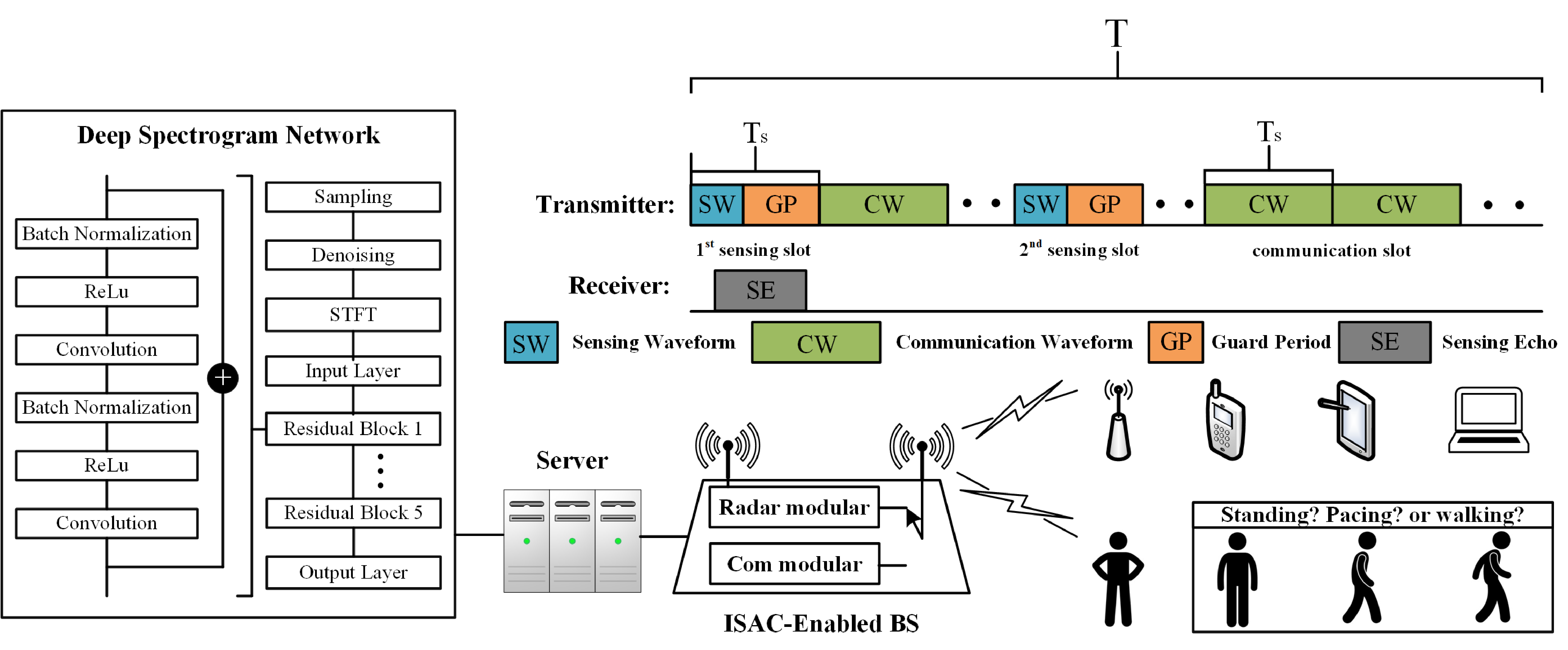}
	\caption{Illustration of the ISAC system model.}
	\label{frame}
\end{figure*}

In this paper, we would like to shed some light on the above open issues. 
In our preliminary work \cite{SPAWC}, a data-assisted hybrid channel model was proposed to simulate the received sensing signals. 
In this paper, we shall extend this work and propose a complete simulation-driven analysis framework, 
namely Simulation-Driven Performance Predictor and oPtimizer (SDP3), 
to quantify the tradeoff between recognition accuracy and communication throughput, which consists of the following three components as illustrated in Fig. \ref{SDP3_diagram}.
\begin{itemize}

	\item[1)]
    The SDP3 data simulator mimicks experimental datasets in a virtual environment consisting of human target, 
	communication receivers, and static objects (e.g., walls), such that the experimental costs can be saved. 

	\item[2)]
    The SDP3 performance predictor trains a deep spectrogram network (DSN) with the above dataset for motion recognition, 
	tests the recognition accuracy, and approximates the relation between recognition accuracy and sensing duration with power function regression. 

	\item[3)]
    The SDP3 performance optimizer illustrates tradeoff performance between recognition accuracy and communication throughput. 
    It is shown that the derived accuracy-throughput (A-T) region consists of a communication saturation zone, a sensing saturation zone, 
	and a sensing-communication adversarial zone, where the last zone achieves the best balance between sensing and communication. 

\end{itemize}

The rest of the paper is organized as follows. The system model considered in this paper is introduced in Section 2. 
The SDP3 data simulator is presented in Section 3. The procedure of the SDP3 performance predictor is described in Section 4. 
The proposed SDP3 performance optimizer and the sensing and communication tradeoff analysis are presented in Section 5. 
The simulation and experiment results are demonstrated in Section 6. Finally, the conclusion is drawn in Section 7. 



\section{System model}



An SDP3 framework to characterize the communication and human motion sensing tradeoff is proposed in this paper. 
In order to demonstrate the procedure of the framework, 
we consider an time-division-multiple-access-based (TDMA-based) ISAC system located in a conference room as an example, 
where the conference room follows the same specification as that in IEEE 802.11ay \cite{ay}. 

As illustrated in Fig. \ref{frame}, the system consists of
one ISAC-enabled base station (BS), $K$ communication receivers and one human target to be sensed.
Both the radar and communication modulars are implemented at the BS, which are multiplexed in time domain.
In order to facilitate transmission and sensing scheduling, the time is organized by slots, 
including the sensing slots and communication slots. 
The slot duration is $T_s$ and the channel impulse response is assumed to be quasi-static 
in each slot. 
The communication modular is enabled in the communication slots for downlink data transmission, 
and the radar modular is enabled in the sensing slots for the motion recognition of the human target.
The sensing and communication tradeoff is studied by adapting the number of sensing and communication slots in every N slots, 
which are referred to as a scheduling period. 
The numbers of communication and sensing slots in a scheduling period are denoted as $N_c$ and $N_s$, respectively. Thus, $N_c+N_s=N$.
Moreover, in order to better capture the micro-Doppler effect, the sensing slots are not successive. 
Instead, there are $N_m$ communication slots between two neighboring sensing slots.
Large $N_s$ could lead to better resolution of micro-Doppler effect and capture more motion details of human target, 
at the price of lower communication throughput. 
The communication and sensing models are elaborated below, respectively.

\subsection{Communication Model}

One receiver is selected in one communication slot for downlink transmission.
The transmission signal is modulated via orthogonal frequency-division multiplexing (OFDM) with M subcarriers. 
Let $h_{k,j}(t)$ be the channel impulse response from the BS to 
the k-th communication receiver in the j-th slot. 
\footnote{Although there is Doppler effect in the channel due to the motion of sensing target, 
the phase shift raised by the Doppler effect is negligible in one slot. Moreover, 
the propagation paths without Doppler shift is dominant in the overall channel impulse response. 
Hence, it is assumed that the CSI is quasi-static in one slot.} 
The channel gain of the m-th subcarrier from the BS to the k-th receiver can be written as 
\begin{align}
H_{k,j,m} = \sum_{l=0}^{L-1}h_{k,j}\left(t-lT\right)e^{\frac{-j2\pi ml}{M}},~0 \leq m \leq M-1,  \label{H_sub}
\end{align}
where T is the sampling period of OFDM transceiver \cite{OFDM}. 
Then, the received signal of the m-th subcarrier is given by
\begin{align}
Y_{k,j,m} = H_{k,j,m}X_{k,j,m}+Z_{k,j,m},  \label{rt1}
\end{align}
where $X_{k,j,m}$ and $Z_{k,j,m} \sim \mathcal{CN}(0,\sigma_z^2)$ are the transmitted signal and the white Gaussian noise, respectively, and $\sigma_z^2$ is the noise power. 
As a result, the throughput of the k-th receiver in the j-th slot (if this receiver is selected) can be expressed as 
\begin{align}
R_{k,j} &=  \frac{T_s}{T_o}\sum\limits_{m=1}^{M} \log_{2} \left(1 + \gamma_{k,j,m}\right)
, \label{R_k}
\end{align}
where $T_o$ is the duration of one complete OFDM symbol (including the cyclic prefix), 
\begin{align}
\gamma_{k,j,m} = \frac{|H_{k,j,m}|^2P_{k,j,m}}{\sigma_z^2},
\end{align}
and $P_{k,j,m}$ is the power allocated to the m-th subcarrier. Notice that the total power constraint $\sum_{m=1}^{M}P_{k,j,m}\leq P$ should be satisfied 
in the power allocation.

\subsection{Sensing Model}

In each sensing slot, the frequency-modulated continuous wave (FMCW) is broadcasted for human motion detection, followed by a guarding interval.
Let $s(t)$ denote the FMCW and $h_{0,j}(t)$ be the channel impulse response between the radar modular transmitter 
and receiver of the j-th slot (if the j-th slot is sensing slot),
the received signal at the radar modular is
\begin{align}
r_j(t)=h_{0,j}(t)\ast s(t)+n_j(t),  \label{rt}
\end{align}
where $n_j(t)$ is the Gaussian noise with average power $\sigma_z^2$, and $s(t)$ follows the power constraint 
$\frac{1}{T}\int_{-\frac{T}{2}}^{\frac{T}{2}}|s(t)|^2dt=P$.

Based on the received signals of the sensing slots $\{r_j(t)|j~\text{mod}~(N_m+1)=1,~1 \leq j \leq N_s(N_m+1)\}$, the human motion can be recognized via a neural network 
which should have been trained in advance via a dataset collected in the same environment. 
Note it is inefficient to collect dataset of $\{r_j(t)|j~\text{mod}~(N_m+1)=1,~1 \leq j \leq N_s(N_m+1)\}$ via experiments for various environments, 
and it is also difficult to investigate the recognition accuracy of neural network with respect to $N_s$ analytically.
The study of communication and sensing tradeoff is challenging.  
In order to address the above issues, in the following Section 3, the SDP3 data simulator based on a data-assisted hybrid channel (DAHC) model is proposed to generate the datasets of received sensing signals via simulation, 
which has been adopted as a part of the channel model for WLAN (wireless local area network) sensing in IEEE 802.11bf \cite{proposal}. 
Based on it, a Deep Spectrogram Network (DSN) for motion recognition is proposed, and the approximated expression of recognition accuracy versus the number of sensing slots, denoted as $A=\Theta(N_s)$, 
is obtained in Section 4, which is referred to as the SDP3 performance predictor. Finally, the tradeoff between the sensing and 
communication performance is studied via SDP3 performance optimizer in Section 5.

\begin{table*}[!t]
\caption{A Comparison of Existing and Proposed Channel Models}
{

\centering
\begin{tabular}{|c|c|c|c|c|c|c|c|c|}
\hline
\hline
\textbf{Type} & \textbf{Literature} & \textbf{Methodology} 
& \textbf{\begin{tabular}[c]{@{}c@{}}Application \\ Scenario \end{tabular}}
& \textbf{\begin{tabular}[c]{@{}c@{}}Spatial-Temporal \\ Consistency \end{tabular}}
& \textbf{\begin{tabular}[c]{@{}c@{}}Motion \\ Consistency \end{tabular}}
& \textbf{\begin{tabular}[c]{@{}c@{}}Micro \\ Doppler \end{tabular}}
& \textbf{\begin{tabular}[c]{@{}c@{}}Sensing \\ Uncertainty \end{tabular}}
\\ \hline
\multirow{2}{*}{\textbf{\begin{tabular}[c]{@{}c@{}}S\end{tabular}}}                     & \multirow{2}{*}{\cite{schannel}}                         & cluster random    & \multirow{2}{*}{commun.}    & \multirow{2}{*}{L1}     & \multirow{2}{*}{\XSolidBrush}      & \multirow{2}{*}{\XSolidBrush}           & \multirow{2}{*}{\XSolidBrush}
\\ 
					&                        & process            &       &           &              &           &
\\ \hline \multirow{2}{*}{\textbf{\begin{tabular}[c]{@{}c@{}}D\end{tabular}}}  & \multirow{2}{*}{\cite{ray}}             & ray                                                              & \multirow{2}{*}{both}                                                                                                                    & \multirow{2}{*}{L2}                                            & \multirow{2}{*}{\Checkmark}                                     & \multirow{2}{*}{\Checkmark}                                                                   & \multirow{2}{*}{\XSolidBrush}
\\&               & tracing                                                               &                                                      &         &                         &                    &                
\\ \cline{2-8} & \multirow{2}{*}{\cite{primitive}}                                           & primitive                                                            & \multirow{2}{*}{sensing}                                                                                                                  & \multirow{2}{*}{L2}                                                      & \multirow{2}{*}{\Checkmark}                                                                  & \multirow{2}{*}{\Checkmark}                                                                   &     \multirow{2}{*}{\XSolidBrush}                                                                                                                      
\\&               & based                                                               &                                                      &         &                         &                    &                
\\ \hline
\multirow{10}{*}{\textbf{\begin{tabular}[c]{@{}c@{}}Q--D \end{tabular}}}  &
3GPP TR & GBSM                                                         & \multirow{2}{*}{commun.}                                                                                                                    & \multirow{2}{*}{L2}                                                   & \multirow{2}{*}{\XSolidBrush}                                                     & \multirow{2}{*}{\XSolidBrush}                                                                    & \multirow{2}{*}{\XSolidBrush}
\\ 
& 38.901 \cite{3gpp} & based                                                        &                                                             &                                                               &                                                                                               &                                                                 & 
\\\cline{2-8}  & \multirow{2}{*}{QuaDRiGa \cite{quadriga}}       & GBSM                                                               & \multirow{2}{*}{commun.}                                                                                                                        & \multirow{2}{*}{L2}                                                           & \multirow{2}{*}{\XSolidBrush}                                      & \multirow{2}{*}{\XSolidBrush}                                                                    &\multirow{2}{*}{\XSolidBrush}
\\&               & based                                                               &                                                      &         &                         &                    &                
\\ \cline{2-8}	& IEEE 802.11        & Q-D               & \multirow{2}{*}{commun.}                                                                                                                      & \multirow{2}{*}{L2}                                                          & \multirow{2}{*}{\XSolidBrush}                                             & \multirow{2}{*}{\XSolidBrush}                                                                    & \multirow{2}{*}{\XSolidBrush}
\\ 
																						& ay \cite{ay}       & based               &                &                 &             &                  & 
\\ \cline{2-8}
																						& \multirow{2}{*}{METIS \cite{METIS}}              &GBSM, map                                                              & \multirow{2}{*}{commun.}            & \multirow{2}{*}{L2}    & \multirow{2}{*}{\XSolidBrush}    & \multirow{2}{*}{\XSolidBrush}        & \multirow{2}{*}{\XSolidBrush}
\\
																						&              &based hybrid                                                              &                                                              &        &          &            & 
\\ \cline{2-8}
																						& \multirow{2}{*}{\textbf{Proposed}}                   & primitive based                                                               & \multirow{2}{*}{both}             & \multirow{2}{*}{L2}    & \multirow{2}{*}{\Checkmark}  & \multirow{2}{*}{\Checkmark}          & \multirow{2}{*}{\Checkmark}
\\
																						&               & hybrid                                                               &                                                      &         &                         &                    &                 \\ \hline
\hline
\end{tabular}
}
\hspace{1cm}

GBSM means geometry-based stochastic channel model.

``S'' means statistical, ``D'' means deterministic, ``Q--D'' means quasi-deterministic.

``L1'' means large-scale spatial consistency, ``L2'' means large-scale and small-scale spatial consistency.

``\checkmark'' means functionality supported, ``\XSolidBrush" means functionality not supported.
\label{table_channel_model}
\end{table*}

\section{Sdp3 data simulator}\label{sec:DAHC}

In this section, we first summarize the drawbacks of the existing channel models in sensing dataset generation, 
then propose a novel data-assisted hybrid channel (DAHC) model for the puropose of efficient sensing dataset generation. Moreover, 
the human kinematic model for the motion simulation is also introduced. 

\subsection{Preliminaries}

\begin{figure*}[!t]
	\begin{center}
		$\begin{array}{ccc} 
		\epsfxsize=2.12 in \epsffile{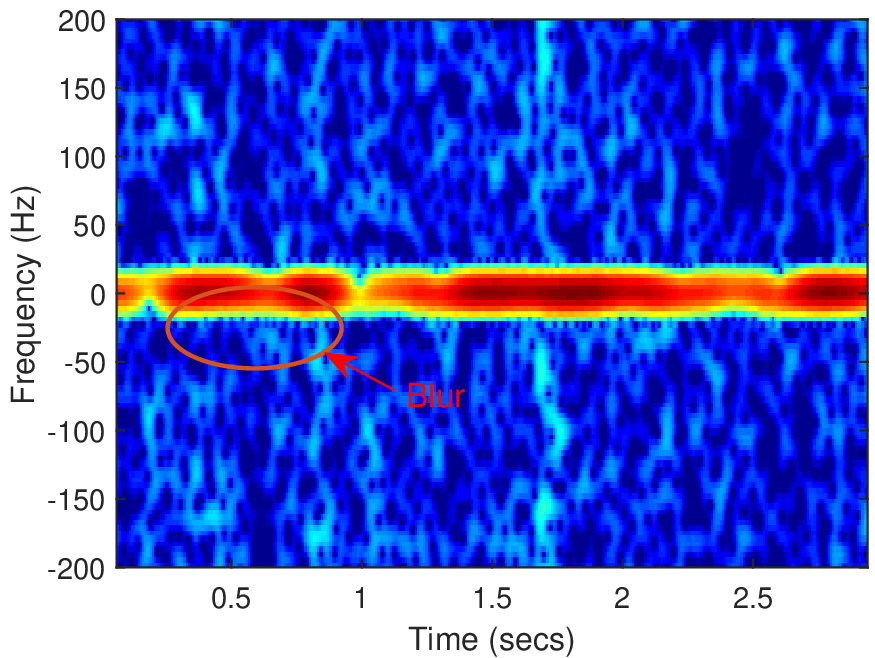} &
		\hspace*{-.2cm}
		\epsfxsize=2.12 in \epsffile{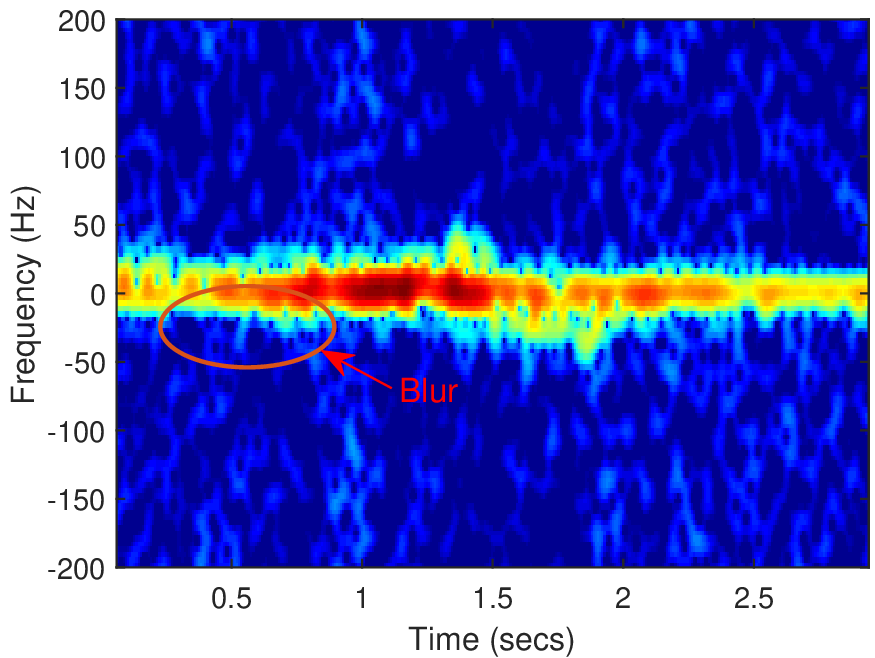} &
		\hspace*{-.2cm}
		\epsfxsize=2.12 in \epsffile{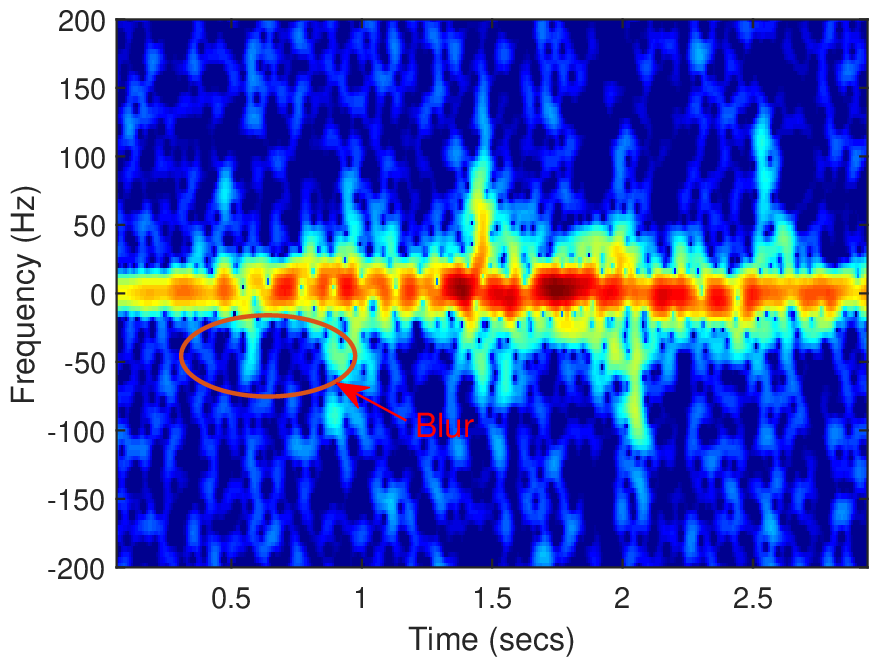} \\ [0.1cm]
		\text{\footnotesize (a) Standing with 3.5GHz in scenario 1} 
		& \text{\footnotesize (b) Pacing with 3.5GHz in scenario 1} 
		& \text{\footnotesize (c) Walking with 3.5GHz in scenario 1} \\ [0.2cm]
		\end{array}$
		$\begin{array}{ccc} 
		\epsfxsize=2.12 in \epsffile{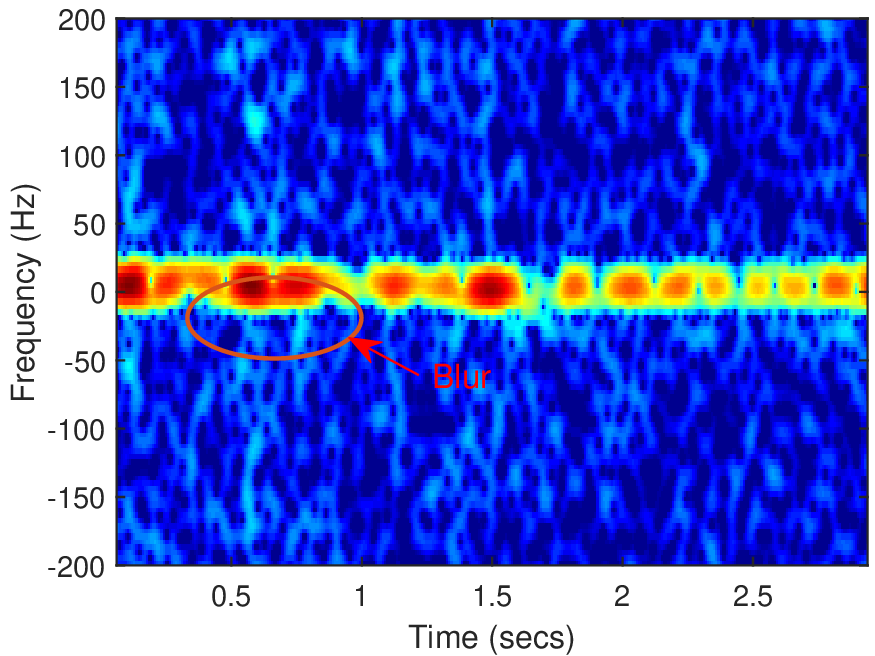} & 
		\hspace*{-.2cm}
		\epsfxsize=2.12 in \epsffile{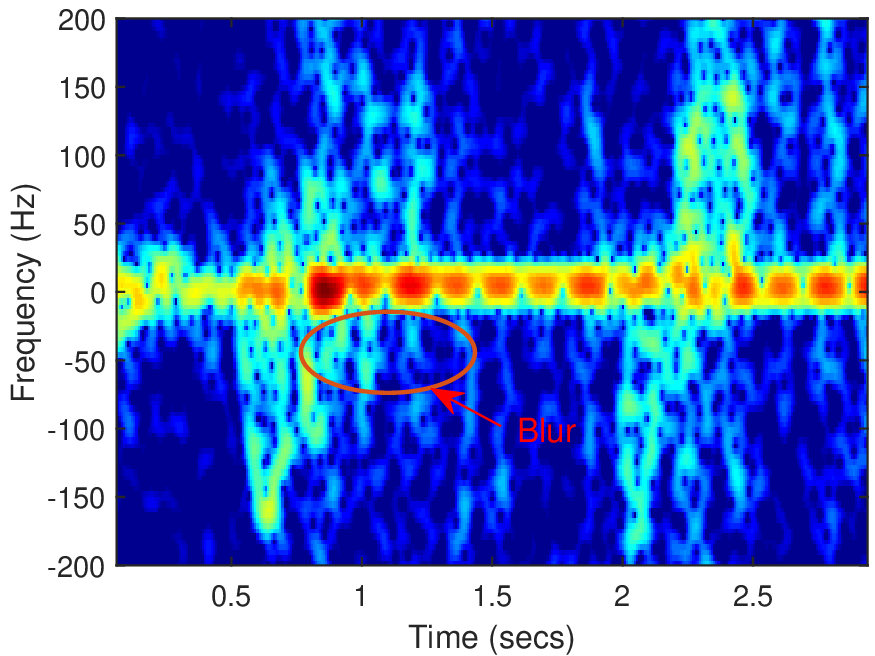} & 
		\hspace*{-.2cm}
		\epsfxsize=2.12 in \epsffile{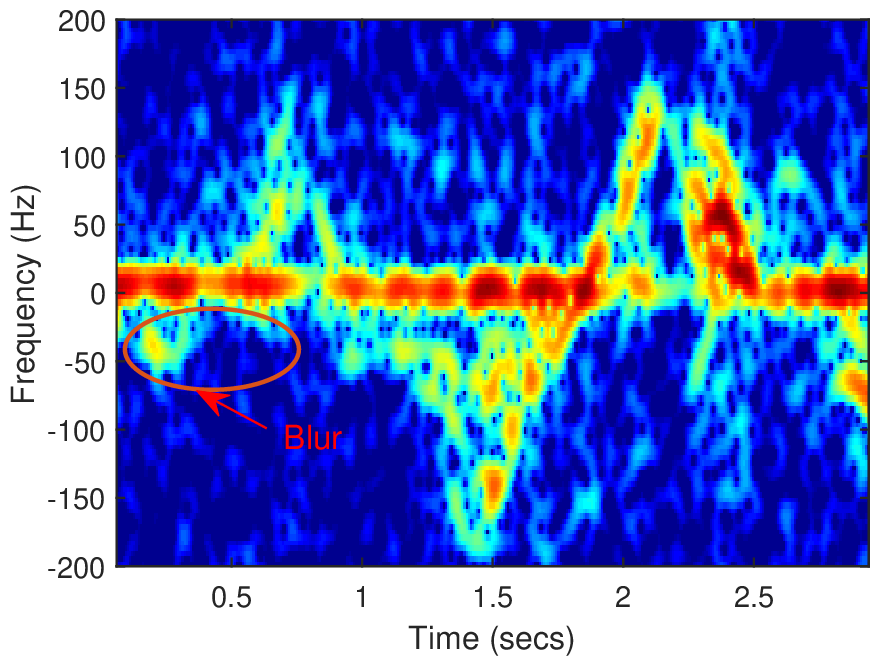} \\ [0.1cm]
		\text{\footnotesize (d) Standing with 60GHz in scenario 2} & 
		\text{\footnotesize (e) Pacing with 60GHz in scenario 2} 
		& \text{\footnotesize (f) Walking with 60GHz in scenario 2} \\ [0cm]
		\end{array}$
		\caption{The experimental results of wireless sensing in scenario 1 and 2.}
		\label{fig:exp_spe}
	\end{center}
\end{figure*}

A wireless channel model for sensing performance evaluation should generate consistent channel impulse response
spatially and temporally so that the receiver can capture the micro-Doppler effect via the received signals in a time interval. 
In fact, the following two kinds of consistency in both spatial and 
temporal domains have been proposed for communication channel model design \cite{METIS}: 
1) large-scale spatial-temporal consistency refers to consistent power fading, delay spreads and angular spreads at 
two close locations and time instances; 
2) small-scale spatial-temporal consistency refers to consistent delays and angles of rays at 
two close locations and time instances. 
Besides them, the channel model for wireless sensing should be composed of the rays consistent with the 
environment and the motions of sensing target. Finally, random interference should also be included to 
model the unpredictable motions except the sensing target.

As summarize in Table \ref{table_channel_model}, current channel models can be categorized 
into statistical models, deterministic models, and quasi-deterministic models.
As an example of the statistical model proposed in \cite{schannel}, the Non-line-of-sight (NLoS) rays between the transmitter and the receiver were 
generated via scattering clusters. The phases, amplitudes and delays of rays were generated via independent distributions.
Lack of the spatial-temporal consistency in small scale, this model cannot simulate the 
micro-Doppler effects due to the non-rigid human motions. 

On the other hand, the quasi-deterministic models could maintain the spatial-temporal consistency in both
large and small scales, which are adopted in many existing industrial standards for communication systems \cite{ay,3gpp,METIS,quadriga}. 
In \cite{3gpp,METIS,quadriga}, the large-scale channel 
parameters, including delay spread, angular spreads, Ricean K-factor
and shadow fading, were computed via ray tracing and the small-scale channel parameters, 
inculding time delays, cluster powers and arrival/departure angles, were obtained using statistical method, 
where the small-scale consistency is maintained via the distribution correlation. 
For example, the delays and angles of rays were generated based on uniform distributions in \cite{3gpp}
where the distribution parameters linearly depend on the correlation distance. 
In \cite{VTC}, a method to calculate the correlation distance according to the large-scale parameters was proposed.
In \cite{ay}, a channel model consisting of both deterministic and random rays was proposed, 
where the deterministic rays were generated via ray tracing to maintain the small-scale consistency. 
However, the modeling of human motions is not considered in the above channel models. 
For example, the human body was simply treated as a blocker of rays in \cite{3gpp} and \cite{ay}.
Thus, the micro-Doppler effect of human motion are not characterized in the above models.

The deterministic channel models such as ray-tracing \cite{ray} have potential to capture information of both environment and human target. 
However, the computation complexity could be huge if the ray tracing method is directly applied on non-rigid human motions.
On the other hand, primitive-based channel model \cite{primitive} is a computationally efficient approximation method for ray tracing, 
where each dynamic object is modeled as an extended target made of multiple point scatters (primitives) distributed along its body (e.g., a human body usually consists of $16$ primitives). 
The received signal from each primitive is computed using the electromagnetic field method in \cite[Sec. 5.8]{moving}, where the radar cross section (RCS) for simple shapes could be obtained from \cite{rcs}.

As a summary, the quasi-deterministic channel models are able to simulate the rays' propagation in the sensing environment, 
and the primitive-based model is able to efficiently simulate the rays from human target. 
The integration of both models could be used for sensing dataset generation. 
Furthermore, due to the complexity and uncertainty of the real sensing scenario, 
it is necessary to keep randomness in the channel impulse response, which is referred to as the \emph{sensing uncertainty} in this paper. 
Specifically, sensing uncertainty could be raised at least by the following three factors:
1) unpredictable reflections from the wall and random scatters,
2) undesired movements of the non-target objects and 3) noise. 
Examples of spectrograms generated from the real experiment are illustrated in Fig. \ref{fig:exp_spe}. 
It can be observed that there are significant blurs aroused by the sensing uncertainty. 
To our best knowledge, the sensing uncertainty has not been not explicitly considered in the existing channel models.

\begin{figure*}[!t]
\centering
		\includegraphics[width=170mm]{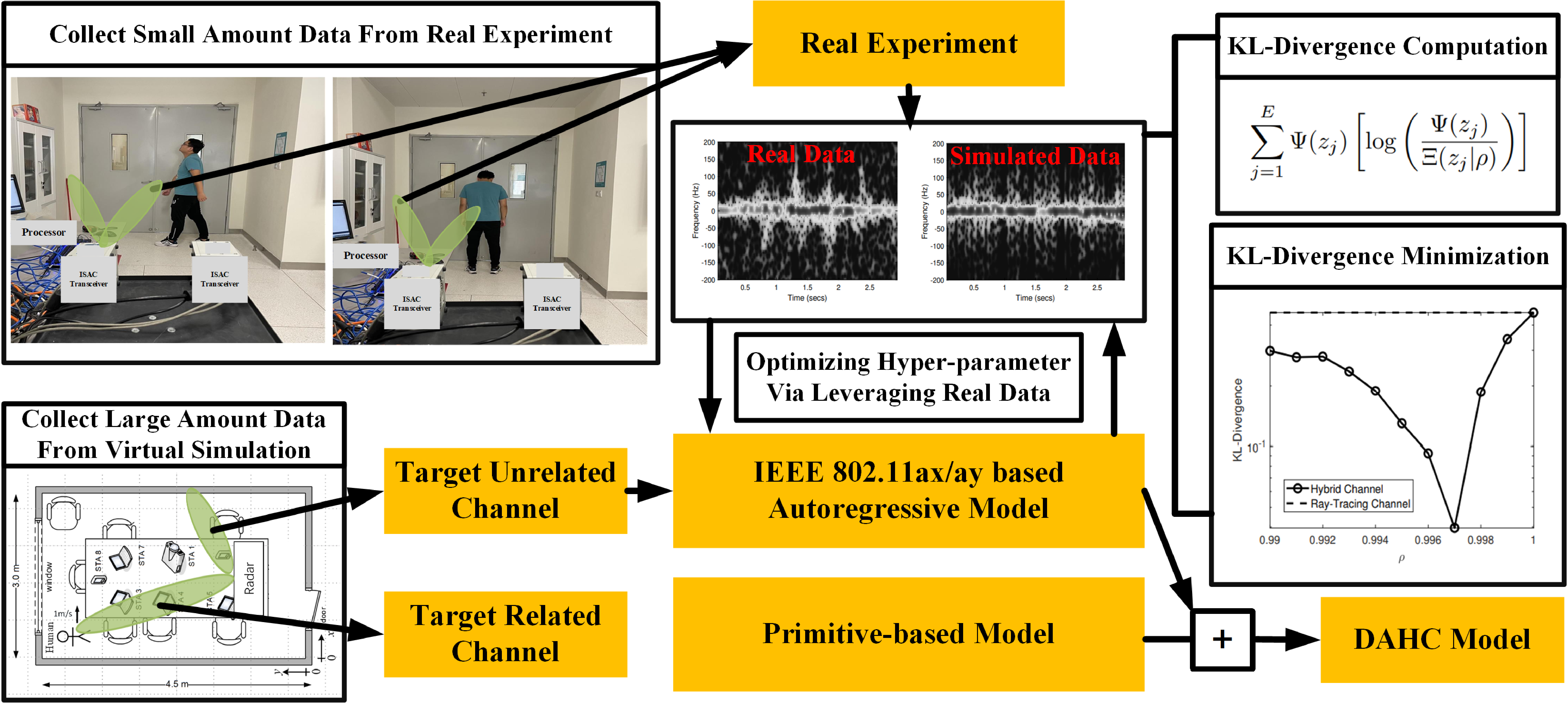}
	\caption{The framework of the proposed DAHC model.}
	\label{framework}
\end{figure*}

\subsection{Channel Modeling} \label{sec:Channel}

In order to keep the consistency with the sensing scenario and model the sensing uncertainty, 
the autoregressive model and quasi-deterministic model are integrated in the 
proposed data-assisted hybrid channel (DAHC) model as illustrated in Fig. \ref{framework}. Specifically,
the channel impulse response from the BS to the k-th receiver 
(the 0-th receiver refers to the co-located BS radar receiver) 
in the j-th time slot can be represented by
\begin{align}
&h_{k,j}(t)=u_{k,j}(t)+v_{k,j}(t), \label{DAHC}
\end{align}
where
\begin{itemize}
\item $u_{k,j}(t)$ is the target-related channel component consisting of the rays reflecting from the human target.
\item $v_{k,j}(t)$ is the target-unrelated channel component, consisting of the rays via the LoS path and the other reflection paths. 
For example, the reflection paths via walls. Due to the self-interference cancellation at the BS radar receiver, we neglect the LoS path in $v_{0,j}(t)$.
\end{itemize}

Similar to \cite{primitive}, the target related channel $u_{k,j}(t)$ is modeled using the following primitive-based method:\\
\begin{align}
u_{k,j}(t)&
=\frac{A}{\sqrt{4\pi}}
\sum_{b=1}^B
  \frac{\sqrt{G_{b,j}}}{D_{b,j}^2}
    \mathrm{exp}\left(-\mathrm{j}\frac{2\pi f_c}{c}2D_{b,j}\right)
    \nonumber\\
    &\quad{}{}
    \times
   \mathrm{exp}\left(\mathrm{j}\varphi_{b}\right)
  \delta\left(t-\frac{2D_{b,j}}{c}\right). \label{uit}
\end{align}
In the above equation, $A$ is the constant related to wave length $\lambda$ and antenna gain $P_t$, which is $\lambda^2 \sqrt{P_t}$ in 
the LoS case; B is the number of primitives; $G_{b,j}$ is the amplitude accounting for radar cross section of the $b$-th primitive in the j-th time slot; 
$D_{b,j}$ is the distance from the $b$-th primitive to the radar in the j-th time slot, $f_c$ is the carrier frequency;
$c$ is the speed of light; $\varphi_{b}$ is the initial phase of the $b$-th ray, 
which follows uniform distribution in $[-\pi,\pi]$. $\delta\left(a\right)$ is the indicator function, whose value is 1 for $a = 0$ and 0 otherwise.

On the other hand, it is not necessary to capture the motions via the primitive-based approach 
in the target-unrelated channel $v_{k,j}(t)$ as that in target-related channel.
Hence, the following autoregressive method is used to model the sensing uncertainty statistically 
\begin{align}
&v_{k,j}(t)=
\left\{
\begin{aligned}
&\Upsilon_1(t)
,~~~~~~~~~~~~~~~~~~~~~~~~~\,\quad\mathrm{if}~j=1
\\ &
\rho v_{k,j-1}(t-T_0)+(1-\rho)\Upsilon_j(t)
,\quad\mathrm{if}~j>1
\end{aligned}
\right.
, \label{vit}
\end{align}
where $\Upsilon_1(t)$ is the target-unrelated channel impulse response of the first slot, and $\rho$ is a hyper-parameter controlling the intensity of sensing uncertainty.
A larger (or smaller) $\rho$ leads to a weaker (or stronger) sensing uncertainty.
In this paper, we adopt the quasi-deterministic channel model in \cite{ay} to generate $\Upsilon_j(t)$, $\forall j$, thus,
\begin{align}
\Upsilon(t)_j&=
\sum_{n=1}^N\sqrt{H_n}\frac{\lambda}{4\pi (D_0+\tau^{\mathrm{cluster}}_{n}c)}
\nonumber\\
&\quad{}{}
\times
\left[\sum_{m=1}^Ma_{n,m}\mathrm{exp}\left(\mathrm{j}\phi_{n,m}\right)
  \delta(t-\tau^{\mathrm{ray}}_{n,m})\right].
\end{align}
In the above expression, N is the number of scattering clusters, 
$H_n$ is the reflection factor for both first-order and second-order reflections; 
$\lambda$ is the wave length;
$\tau^{\mathrm{cluster}}_{n}$ is the $n$-th cluster's time delay (in seconds) obtained from ray-tracing; 
$\tau^{\mathrm{ray}}_{n,m}$, $a_{n,m}$ and $\phi_{n,m}$ are the time delay, 
amplitude and initial phase of $m$-th ray via the $n$-th cluster, which are obtained from Poisson distribution, Rayleigh distribution and uniform distribution respectively. 

\subsection{Data-Assisted Model Calibration}

In order to fit the hyper-parameter $\rho$ of the DAHC to the real uncertainty level in a particular sensing scenario, 
it is necessary to make a measurement in the target environment. Denote the received signal in the real scenario and DAHC model simulation as
$\{r_j(t)|j~\text{mod}~(N_m+1)=1,~1 \leq j \leq N_s(N_m+1)\}$ and $\{\hat{r}_j(t)|j~\text{mod}~(N_m+1)=1,~1 \leq j \leq N_s(N_m+1)\}$, respectively. 
The adaptation of hyper-parameter to real uncertainty level is elaborated below.

The spectrograms of both measured and simulated signals are first obtained according to Section \ref{sec:predictor}. 
The Doppler frequency strength versus time and frequency of both spectrograms are 
quantized into F levels. The probability mass function (PMF) of Doppler frequency strength for both spectrograms can then be obtained, 
denoted as $\Psi$ and $\Xi$ respectively.
Finally, the proper hyper-parameter $\rho$ is the one minimizing the Kullback-Leibler (KL) divergence as follows.
\begin{align}
\mathop{\mathrm{min}}_{\rho}\quad&\sum\limits_{f=1}^{F}\Psi(z_f)\left[\mathrm{log}\left(\frac{\Psi(z_f)}{\Xi(z_f|\rho)}\right)\right],
\nonumber\\
\mathrm{s.t.}\quad
&0\leq\rho\leq 1,
\label{fitting}
\end{align}
where $z_f$ denotes the signal strength of the f-th level. The above problem can be solved by one-dimensional search.
In practice, $\rho$ varies in different scenarios and we can store the values of $\rho$ for typical scenarios in a look-up table.

\subsection{Human Kinematic Modeling} \label{sec:Motion}

The human target is represented by 16 primitives in the proposed DAHC model. 
In order to model the motion of the 16 primitives, a global human model, 
namely the Thalmann model \cite{Thalmann}, is adopted. In the Thalmann model, the human body is represented as 
a series of 16 segments, which corresponds to the primitives of the proposed channel model. 
Based on the biomechanical experimental data, the motions of all the segments, including the their positions and orientations of all time slots, 
are obtained for different categories human motions. 

However, this global human model averages out the personification in the same category of motion, 
losing the diveristy in motion's dataset generation. 
In the future work, the motion capture methods from the the areas of Graphics \cite{Mocap} could be exploited to simulate the primitives' motion with personification.

\section{Sdp3 performance predictor} \label{sec:predictor}

The SDP3 performance predictor is trained to predict the sensing performance versus the sensing resource. Specifically, 
with the motion sensing dataset generated by the SDP3 data simulator, 
a deep spectrogram network (DSN) is trained for motion recognition, then 
the approximate expression of the motion detection accuracy versus the number of sensing slots $N_s$ is derived.

\subsection{Deep Spectrogram Network} \label{subsec:dsn}

Let $C$ and $\mathcal{C}=\left\{1,\dots,C\right\}$ be the number of human motion categories and the set of human motion categories respectively, 
the input of DSN is the signals $\{r_j(t)|j~\text{mod}~(N_m+1)=1,~1 \leq j \leq N_s(N_m+1)\}$ and the output is the index of human motion category $\widehat{c} \in \mathcal{C}$.
In DSN, the spectrogram of input signals is first generated via data cleaning and transformation, 
followed by model training (in the training phase) or inference (in the inference phase), as shown in Fig. \ref{frame}.

\textbf{Data Cleaning}.
In this step, the received signal $r_j(t)$ within the sweep duration $t \in [0,T_{sw}]$ is sampled with a frequency $f_s$. 
Let $\mathbf{x}_j = [r_j(\frac{1}{f_s})~r_j(\frac{2}{f_s})\dots r_j(T_{sw})]^T \in \mathbb{C}^{L}$ 
be the vector of samples in the j-th slot, where $L = T_{sw}f_s$, and 
$\mathbf{X} = \left[\mathbf{x}_1~\mathbf{x}_2~\cdots~\mathbf{x}_{N_s}\right]\in\mathbb{C}^{L\times N_s}$ be the aggregation of sample vectors of all sensing slots.
The matrix $\mathbf{X}$ is a superposed signal consisting of the desired signals reflected from the target and undesired signals reflected from walls (or other objects).
To extract the desired information, we first dechirp the sampled signal $\mathbf{X}$ as follows 
\begin{align}
\widetilde{\mathbf{X}}
=\left(\left[\widetilde{\mathbf{s}}\odot\mathbf{x}_1^{*}~\widetilde{\mathbf{s}}\odot\mathbf{x}_2^{*}~\cdots~ \widetilde{\mathbf{s}}\odot\mathbf{x}_{N_s}^{*}\right]\right)^{*},  \label{X}
\end{align}
where $\widetilde{\mathbf{s}} = [s(\frac{1}{f_s})~s(\frac{2}{f_s})\dots s(T_{sw})]^T$, $\mathbf{x}_1^{*}$ denotes the conjugate of $\mathbf{x}_1$, and $\odot$ denotes the Hadamard product. 
Then the singular value decomposition (SVD) is applied to $\widetilde{\mathbf{X}}$, yielding $\widetilde{\mathbf{X}}=\sum_{i=1}^Da_i\mathbf{u}_i\mathbf{v}_i^H$, where $D$ is the rank of $\widetilde{\mathbf{X}}$, $a_i$ is the i-th largest singular value, $\mathbf{u}_i$ and $\mathbf{v}_i$ are the $i$-th left-singular vector and right-singular vector respectively.
Removing the first $d-1$ components, which represent the undesired signal paths, the denoised signal matrix is 
\begin{align}
\mathbf{Y}=\sum_{i=d}^Da_i\mathbf{u}_i\mathbf{v}_i^H. \label{Y}
\end{align}

\textbf{Data Transformation}.
In this step, the Short Time Fourier transform (STFT) \cite{vcchen} is applied on $\mathbf{Y}$.
We first define the sliding window function as 
\begin{align}
w_{\beta}\left[n\right] =& \frac{I_0\bigg(\beta\sqrt{1-\left(\frac{n-(N_w-1)/2}{((N_w-1)/2)}\right)^2}\bigg)}{I_{0}\left(\beta\right)}, 0\leq n\leq N_w-1, \nonumber
\end{align}
where $I_0$ is the zeroth-order modified Bessel function of the first kind and $N_w$ is the length of sliding window. 
Let $\mathbf{y}=\mathbf{1}^T\mathbf{Y}$, the STFT of $\mathbf{y}$ at time $k$ and the frequency $f$ 
with the sliding window function $w$ can be expressed as
\begin{align}
z\left[k,f\right] =& \sum_{k^{'}=-\infty}^{+\infty} y\left[k^{'}\right]w_{\beta}\left[k^{'}-k\right]\exp(-\mathrm{j}2\pi k^{'}f/N), \nonumber\\
&k \in \left\{0,\dots,\left\lfloor \frac{N_s-N_w}{N_w-N_{overlap}} \right\rfloor\right\}*\left(N_w-N_{overlap}\right), \nonumber\\
&f \in \left\{-\frac{N_{fft}}{2}+1,\dots,\frac{N_{fft}}{2}\right\}, \label{z}
\end{align}
where $N_{overlap}$ and $N_{fft}$ specify the number of overlap samples between adjoining STFT windows and the number of frequency points respectively. 
Then the spectrogram of sensing signal $\{\mathbf{x}_j|j~\text{mod}~(N_m+1)=1,~1 \leq j \leq N\}$ could be illustrated via $z\left[k,f\right]$.

\textbf{Model Training and Inference}.
To classify the motions, the ResNet-32 \cite{ResNet} is adopted as the backbone for the feature extraction. 
It consists of one input layer, five identical residual blocks and one output layer as illustrated in Fig. \ref{frame}. 
The input layer consists of a convolution layer and a pooling layer, and 
the output layer consists of a global average pooling layer and a fully-connected layer with softmax as the activation function. 
Each residual block consists of six layers: a batch normalization layer, a ReLu activation layer, a convolution layer, a batch normalization layer, a ReLu activation layer, and a convolution layer.
The input of the ResNet is the spectrogram, and the output is the index of the estimated human motion category $\widehat{c}$.

\begin{table}[!t]
\renewcommand\arraystretch{1.1}
\caption{Candidate Parameteric Learning Curve Models}
{
\begin{center}
\begin{tabular}{|c|c|c|}
\hline
\hline
\textbf{Name} & \textbf{Experssion} & \textbf{Tuning Parameters} \\\hline
vapor & \multirow{2}{*}{$\mathrm{exp(\alpha+\beta/N_s)}$} & 
\multirow{2}{*}{$\mathrm{\alpha,\beta}$} \\
pressure & & \\\hline
pow$_3$   & $\mathrm{\gamma-\alpha N_s^{-\beta}}$ & $\mathrm{\alpha,\beta,\gamma}$ \\\hline
log power   & $\mathrm{\alpha/(1+(N_s/e^{\beta})^{\gamma})}$ & $\mathrm{\alpha,\beta,\gamma}$ \\\hline
exp$_4$  & $\mathrm{\gamma-e^{-\alpha N_s^{\epsilon}+\beta}}$ & $\mathrm{\alpha,\beta,\gamma,\epsilon}$ \\\hline
log log  & \multirow{2}{*}{$\mathrm{log(\alpha log(N_s)+\beta)}$} & 
\multirow{2}{*}{$\mathrm{\alpha,\beta}$} \\
linear & & \\\hline
ilog2  & $\mathrm{\beta-\alpha/log(N_s)}$ & $\mathrm{\alpha,\beta}$ \\\hline
pow$_4$  & $\mathrm{\gamma-(\alpha N_s+\beta)^{\epsilon}}$ & $\mathrm{\alpha,\beta,\gamma,\epsilon}$ \\\hline
\hline
\end{tabular}
\end{center}
}
\hspace{1cm}
\label{table_learning_model}
\end{table}

\subsection{Motion Recognition Accuracy Model} \label{subsec:accuracy_model}

It is difficult to directly derive the analytical relationship between the motion recognition accuracy and 
the number of sensing slots, which is denoted as $A = \Theta(N_s)$.
This is because there is no analytical expression to quantify the learning performance of ResNet--$32$.

To address the above challenge, a promising solution is the performance regression approach proposed in
\cite{lcpa,lcwra,curve1,curve2}. Specifically, it is observed that the recognition accuracy $A=\Theta(N_s)$ is a nonlinear function of $N_s$ satisfying the following properties:
\begin{itemize}
\item[(i)] $\Theta(N_s)$ is a monotonically increasing function of $N_s$;

\item[(ii)] As $N_s$ increases, the magnitude of the partial derivative $|\partial \Theta/\partial N_s|$ would gradually decrease and become zero when $N_s$ is sufficiently large, meaning that increasing the number of sensing time slots will not help wireless sensing performance at large $N_s$. 

\end{itemize}
Hence the candidate parametric learning curve models to approximate the $\Theta(N_s)$ are shown in Table \ref{table_learning_model} (as proposed in \cite{curve1}), where $\alpha,\beta,\gamma,\epsilon$ are tuning parameters.

In order to determine the tuning parameters, we first generate the dataset of sensing signals 
$\{r_j(t)|j~\text{mod}~(N_m+1)=1,~1 \leq j \leq N_s(N_m+1)\}$ for Q different numbers of sensing slots, denoted as 
$\{N_s^{(i)}|~1 \leq i \leq Q\}$. After the training and inferencing with DSN, the corresponding recognition accuracies are denoted as 
$\{A^{(i)}|~1 \leq i \leq Q\}$. Then the parameters $\alpha,\beta,\gamma,\epsilon$ can be calculated via the following least squares fitting,
\begin{align}
\mathop{\mathrm{min}}_{\alpha,\beta,\gamma,\epsilon}\quad&\frac{1}{Q}\mathop{\sum}_{i=1}^Q\Big|\Theta(N_s^{(i)})-A^{(i)}\Big|^2.
\label{fitting1}
\end{align}
The above problem can be solved by brute-force search, or gradient descent method. 

\section{SDP3 performance optimizer and tradeoff analysis}


The SDP3 performance optimizer, which investigates the accuracy-throughput (A-T) region of the ISAC system, 
is presented in this section. Denote $N_{c,k}$ as the number of time slots assigned for the k-th receiver, 
the throughput of k-th receiver in the whole scheduling period $R_k$ can be approximated as
\begin{align}
R_{k} &\approx  \frac{N_{c,k}T_s}{T_o}\sum\limits_{m=1}^{M} \log_{2} \left(1 + \gamma_{k,m}\right), \label{R_k_2}
\end{align}
where $\gamma_{k,m}=\mathbb{E}\left[\gamma_{k,j,m}\right] = \frac{\mathbb{E}\left[|H_{k,m}|^2P_{k,j,m}\right]}{\sigma_z^2}$. 
In (\ref{R_k_2}), we use $\gamma_{k,m}$ to approximate the instantaneous SNR. This is because the Doppler effect raised by human motion 
does not dominate the channel gain.
Hence the worst communication throughput among all receivers is
\begin{align}
R = \mathop{\mathrm{min}}_{k=1,\cdots,K}
\frac{N_{c,k}T_s}{T_o}\sum\limits_{m=1}^{M} \log_{2} \left(1 + \gamma_{k,m}\right). \label{R}
\end{align}

In order to characterize the A-T region, we first define the following weighted summation of recognition accuracy $A$ and worst communication throughput $R$.
\begin{align}
f(w_A,w_R) = w_A A+w_R R, \label{Obj}
\end{align}
where $w_A$ and $w_R$ are the weights. Given a pair of weight $\left(w_A,~w_R\right)$, 
the optimal $A^{*}$ and $R^{*}$ maximizing the objective $f(w_A,w_R)$ could be obtained via the following optimization problem. 
\begin{align}
\mathcal{P}:
(A^{*},R^{*})=&\mathop{\mathrm{argmax}}_{\substack{A,R,N_s,\mathbf{N_c}}}
~~f(w_A,w_R),
\nonumber\\
\mathrm{s.t.}~~~~
&\text{constraint in }\eqref{R} \nonumber\\
&\sum\limits_{m=1}^{M}P_{k,m}\leq P,~\forall k \nonumber\\
&N_s+\sum\limits_{k=1}^{K}N_{c,k}=N.
\label{Problem}
\end{align}

\begin{figure*}[!t]
	\centering
\includegraphics[width=170mm]{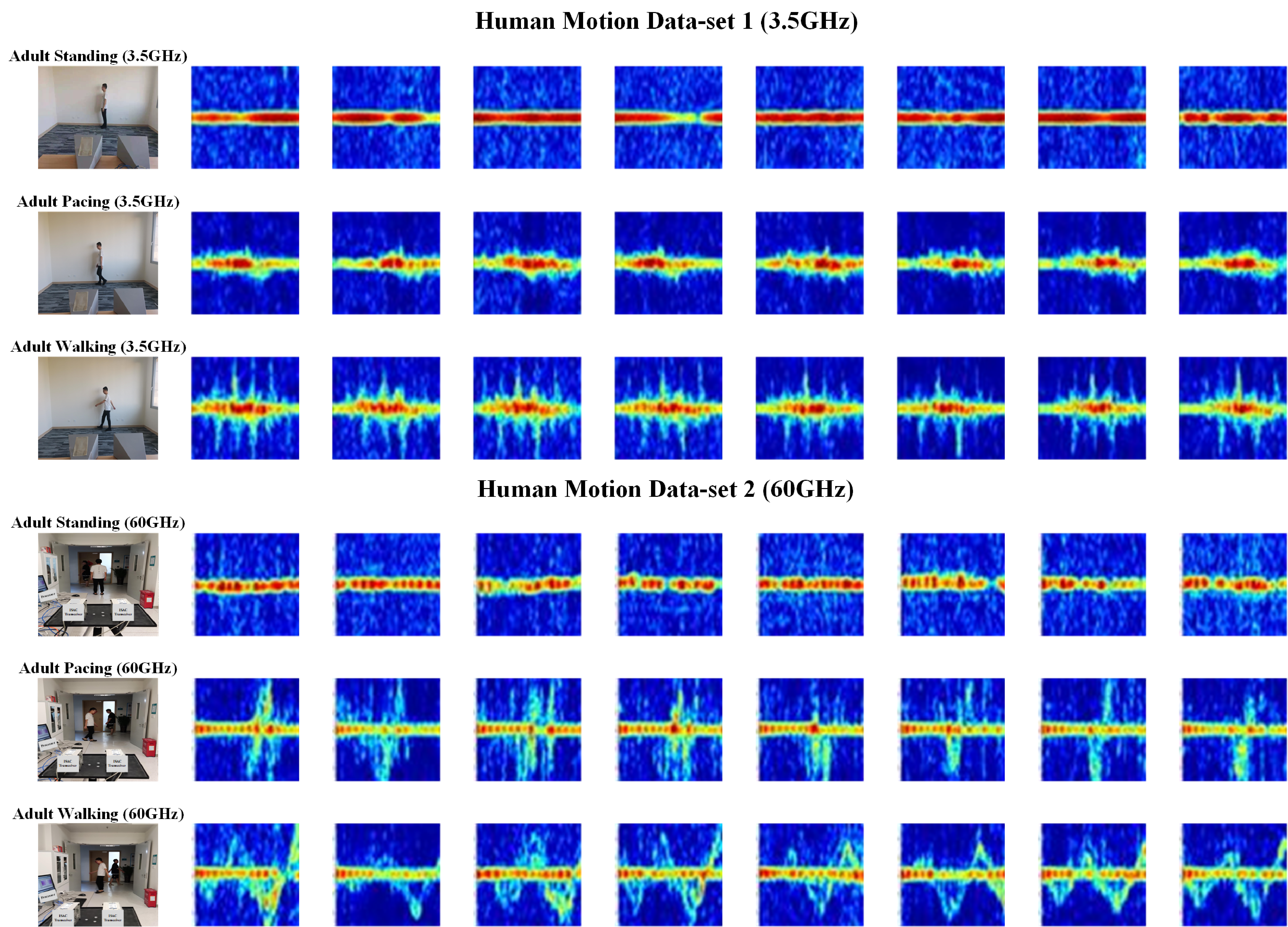}
		\caption{The human motion datasets obtained by sensing an adult target in a conference room, where the carrier frequencies are $3.5\,\mathrm{GHz}$ (upper side) and $60\,\mathrm{GHz}$ (lower side) respectively. 
		The scenarios with carrier frequencies 3.5GHz and 60GHz are referred to as the scenario 1 and 2 respectively. 
		The height of the adult is 1.75m. The following three motions are tested: standing, pacing and walking. 
		The FMCW is with $10\,\mathrm{MHz}$ bandwidth and $100\,\mathrm{\mu s}$ sweep time.}
		\label{Real_Data_Set} 
\end{figure*}

\begin{figure*}[!t]
	\begin{center}
		$\begin{array}{ccc} 
		\epsfxsize=2.05 in \epsffile{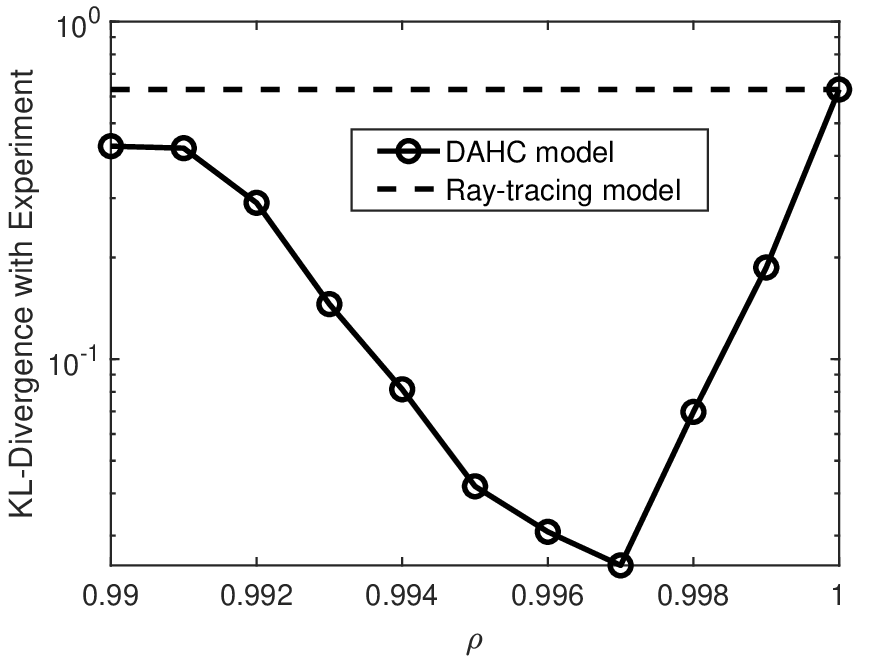} 
		\hspace{.55cm}
		\epsfxsize=2.05 in \epsffile{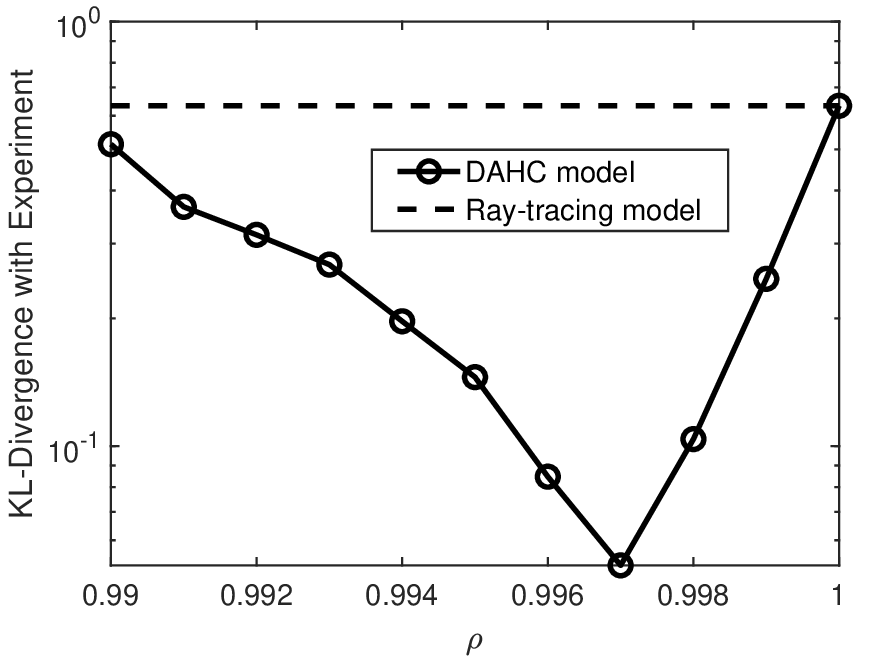} 
		\hspace{.55cm}
		\epsfxsize=2.05 in \epsffile{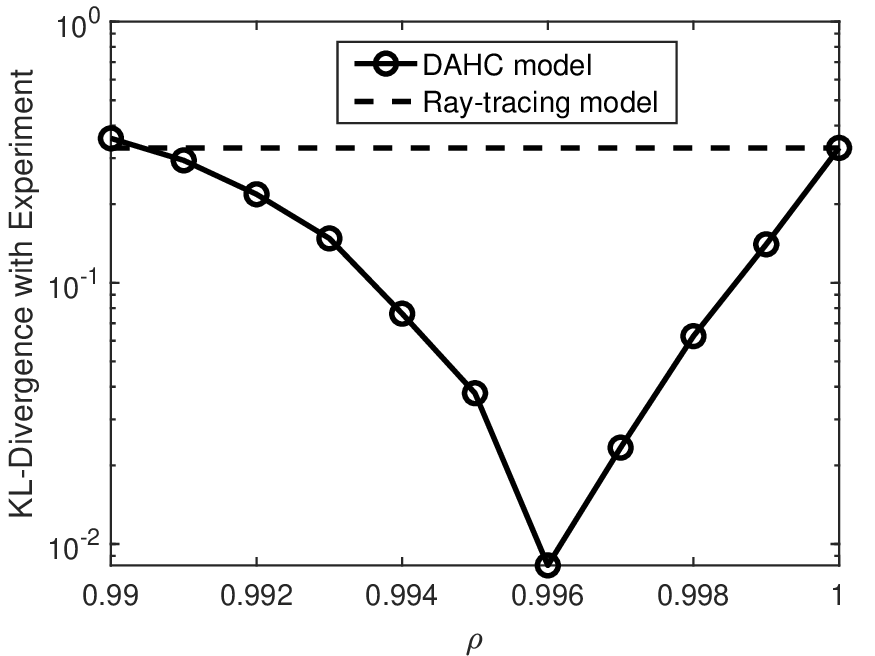} \\ [0.1cm]
		\text{\footnotesize (a) KL divergence on 3.5GHz in scenario 1 } 
		\text{\footnotesize (b) KL divergence on 60GHz in scenario 1 } 
		\text{\footnotesize (c) KL divergence on 60GHz in scenario 2} \\ [0.2cm]
		\end{array}$
		$\begin{array}{ccc} 
		\epsfxsize=2.05 in \epsffile{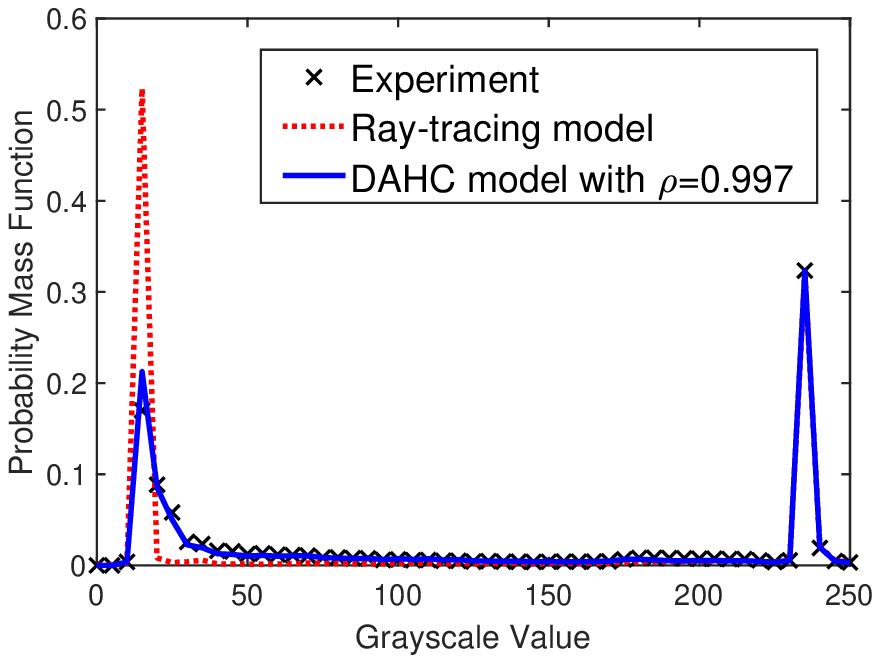} 
		\hspace{.55cm}
		\epsfxsize=2.05 in \epsffile{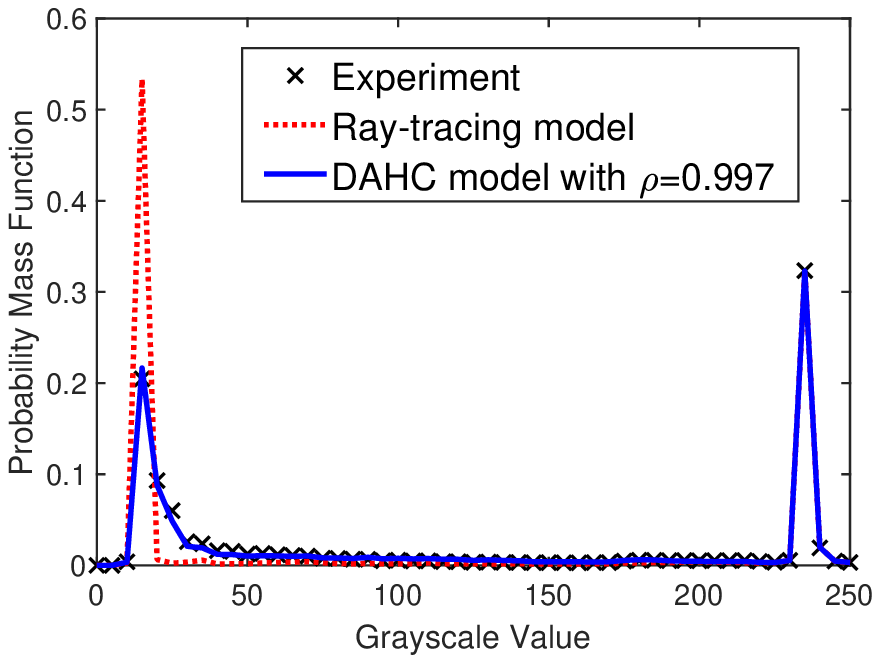} 
		\hspace{.55cm}
		\epsfxsize=2.05 in \epsffile{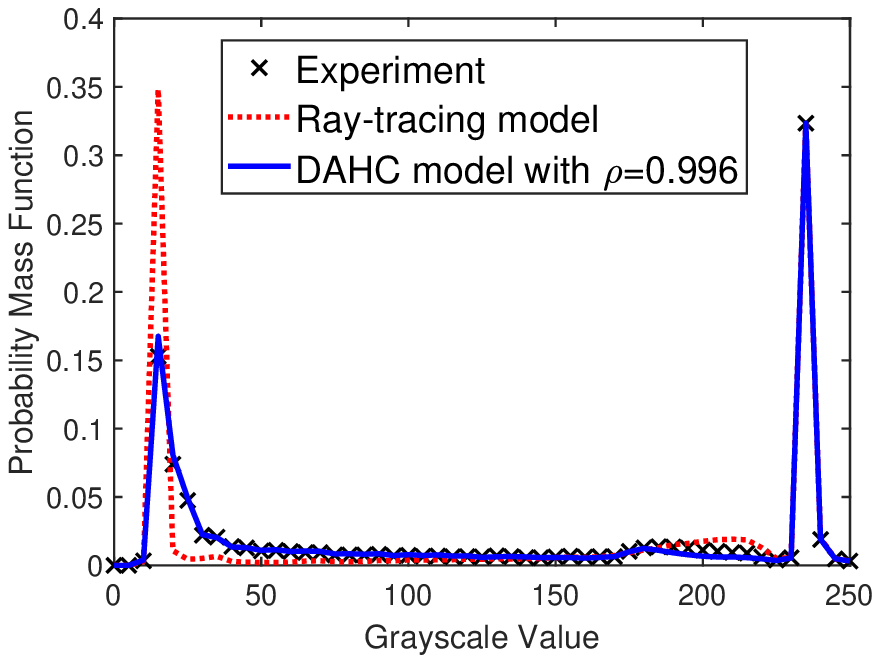} \\ [0.1cm]
		\text{\footnotesize (d) PMF comparison on 3.5GHz in scenario 1 } 
		\text{\footnotesize (e) PMF comparison on 60GHz in scenario 1 } 
		\text{\footnotesize (f) PMF comparison on 60GHz in scenario 2} \\ [0cm]
		\end{array}$
		\caption{The calibration results of DAHC model for walking in scenario 1 and 2.}
		\label{fig:KLD}
	\end{center}
\end{figure*}

Hence, the achievable A-T region can be characterized by $\{(A,R)=\text{argmax}~f(w_A,w_R)|w_A>0,~w_R>0\}$. Moreover, 
the closed-form relationship between the recognition accuracy and communication throughput on the Pareto boundary is 
elaborated in the following theorem.
\begin{theorem}
Denote $\Theta^{-1}$ as the inverse function of $\Theta$, let (A*, R*) be a point on the Pareto boundary, then
\begin{align}
T_s\Theta^{-1}(A^*)+
\left[\mathop{\sum}_{k=1}^K\frac{T_o}{\sum\limits_{m=1 \atop \gamma_{k,m}\geq \gamma_{k,0}}^{N} \log_{2} \left(\frac{\gamma_{k,m}}{\gamma_{k,0}}\right)}\right]R^*
=NT_s.
\end{align}
\end{theorem}
\begin{proof}
Please refer to Appendix A.
\end{proof}

\section{Simulation and experiment} \label{sec:Result}

In this section, we verify the proposed SDP3 framework via both simulation and experiment. 
Specifically, a conference room with size $(3\,\mathrm{m}$, $4.5\,\mathrm{m}$, $3\,\mathrm{m})$ (i.e., length, width, height)
is considered, where the lower left conner of the room is the origin of the coordinates and the radar is located as $(1.5,1,1)$. 
The sensing target is either an adult or a child at the location $(3,4.2,0)$ initially. 
The motions to be classified include the child/adult standing, child walking, child pacing, adult walking and adult pacing. 
In order to keep the consistency between the simulation and experiment, the experiment is conducted in the rooms with similar configuration. 
In both simulation and experiment, two carrier frequencies $f_c=3.5\,\mathrm{GHz}$ and $60\,\mathrm{GHz}$ are considered. 
The total transmit power is $P=1\,\mathrm{W}$ for both sensing and communication and the bandwidth is $B=10\,\mathrm{MHz}$.
A directional antenna with gain $P_t=25\,\mathrm{dB}$ is used for sensing, which is directed to the sensing target. 
An omni-directional antenna is used for communication. The datasets of received sensing signals are generated via both experiment and simulation. 

Specifically, the experiments are conducted in two scenarios and the human motion datasets are also generated shown in Fig. \ref{Real_Data_Set}: (1) there is only one person in the room, who is the sensing target; 
(2) there are two persons in the room, wherein one of them is the sensing target and the other one raises sensing interference. 
The Scenario 1 is sensed on both 3.5GHz and 60GHz band, and the Scenario 2 is sensed on 60GHz band only. In fact, the Scenario 2 is common in practice. 
It refers to the situation with stronger sensing uncertainty. 
In the simulation dataset, we generate the samples of received sensing signal via both SDP3 data simulator and ray-tracing model, such that their performances can be compared. 
In both approaches, the adult and child have $B=16$ primitives respectively. In the SDP3 data simulator, the IEEE 802.11ax/ay channels \cite{ax,ay} are adopted to generate the component $v_{k,j}(t)$ in \eqref{DAHC}. 
In both simulation and experiment datasets, 200 samples are generated for each motion.

\subsection{KL Divergence}

In this part, the calibration of hyper-parameter $\rho$ for the motion of adult walking, as elaborated in Section \ref{sec:DAHC}, 
is first demonstrated, which verifies the existence of sensing uncertainty. 

\begin{figure*}[!t]
	\begin{center}
		$\begin{array}{cc} 
		\epsfysize=2.4 in \epsffile{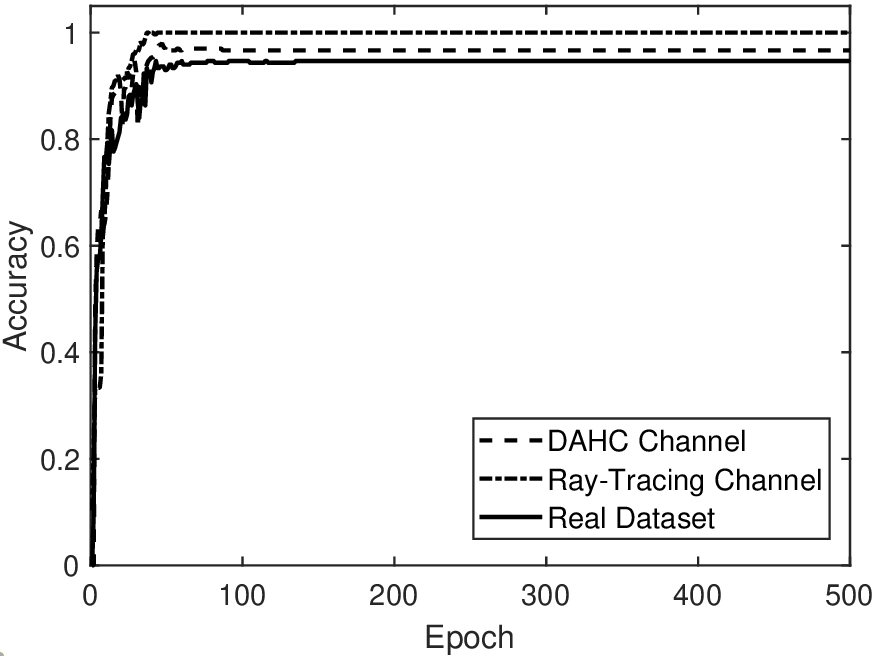} &
		\hspace*{-.2cm}
		\epsfysize=2.4 in \epsffile{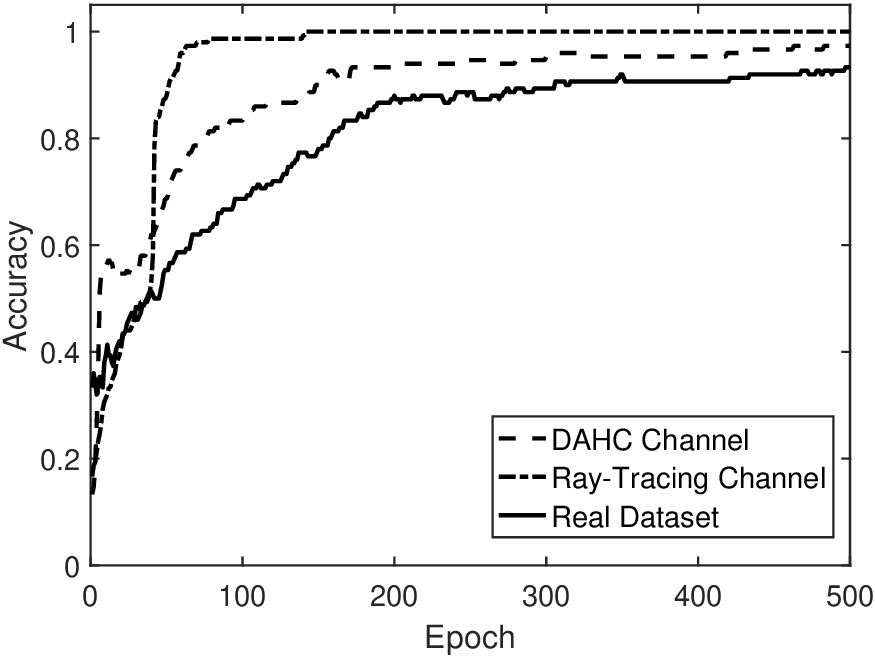} \\ [0.1cm]
		\text{\footnotesize (a) 3.5GHz} & \text{\footnotesize (b) 60GHz} \\ [0cm]
		\end{array}$
		\caption{Comparison of recognition accuracy among the datasets generated by real experiments, DAHC channel model and ray-tracing model 
		with different carrier frequency.}
		\label{fig:Comp}
	\end{center}
\end{figure*}

\begin{figure*}[!t]
	\begin{center}
		$\begin{array}{ccc} 
		\epsfxsize=2.3 in \epsffile{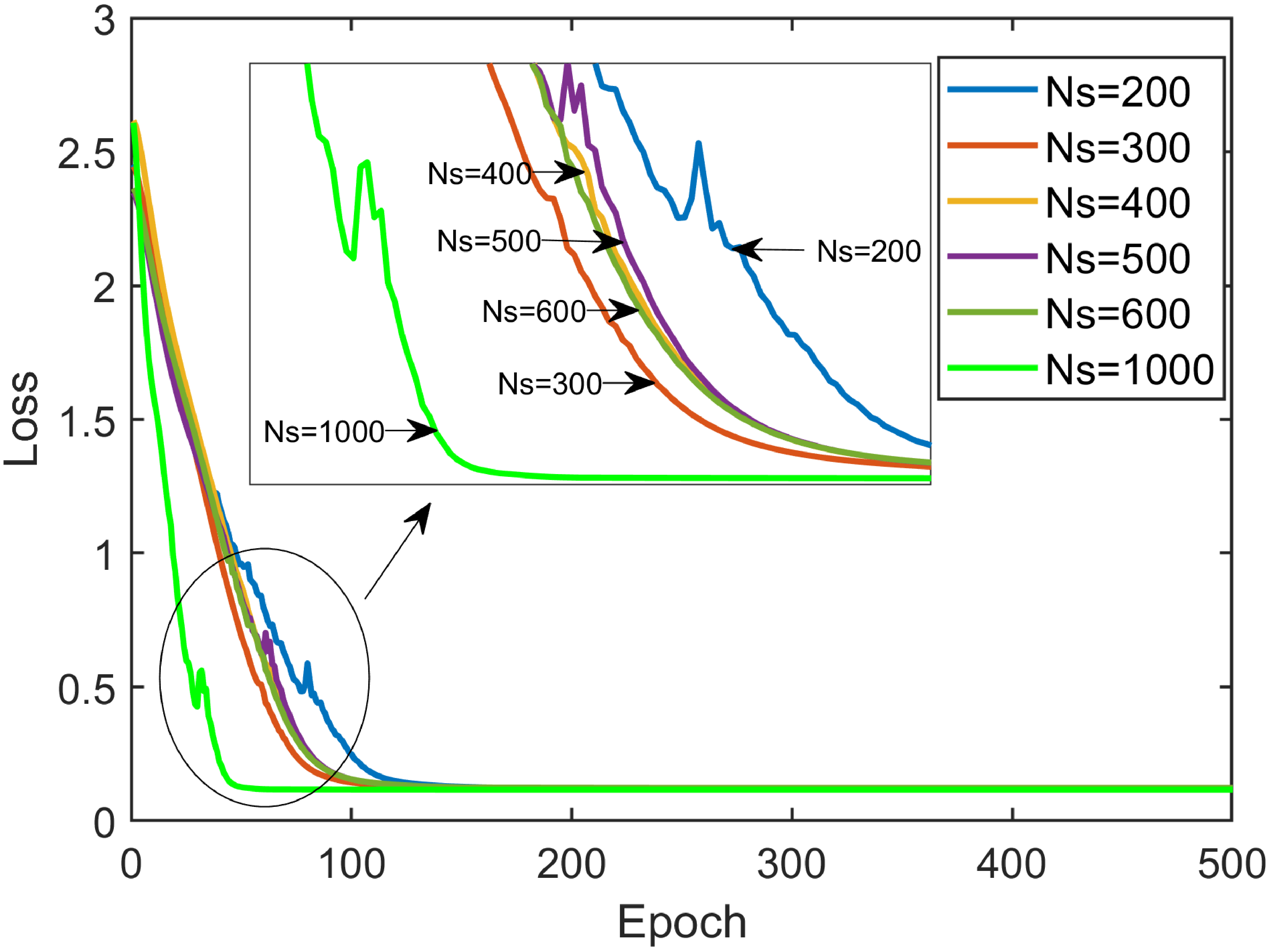} &
		\hspace*{-.2cm}
		\epsfxsize=2.3 in \epsffile{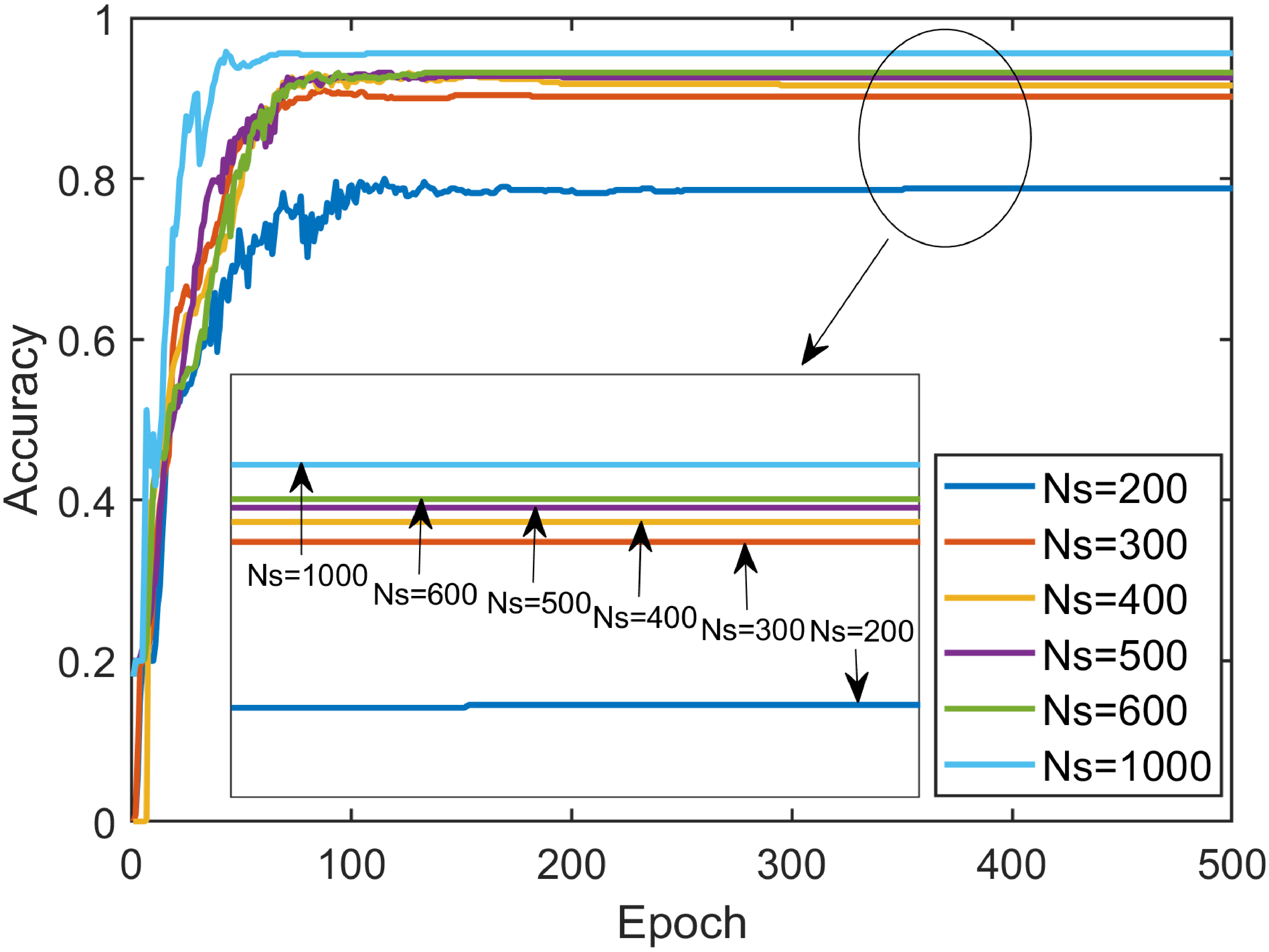} &
		\hspace*{-.2cm}
		\epsfxsize=2.3 in \epsffile{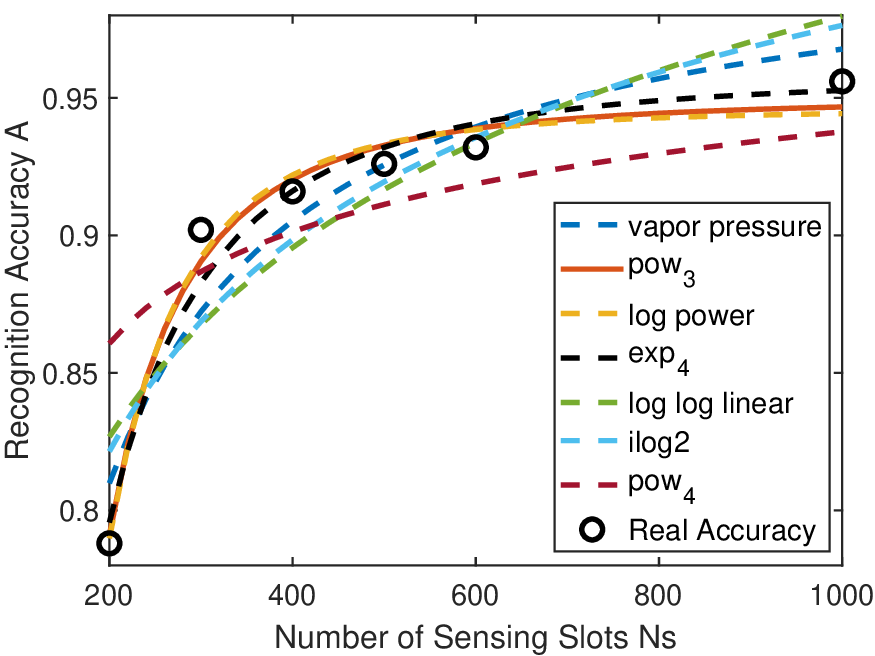} \\ [0.1cm]
		\text{\footnotesize (a) Training loss} & 
		\text{\footnotesize (b) Recognition accuracy} & 
		\text{\footnotesize (c) Learning accuracy approximation} \\ [0cm]
		\end{array}$
		\caption{The motion recognition performance of DSN and approximation result.}
		\label{fig:Training}
	\end{center}
\end{figure*}

Specifically, the KL divergence between the experimental and SDP3-simulated data samples versus the hyper-parameter $\rho$ is shown in Fig.~\ref{fig:KLD}(a-c) for different carrier frequencies and scenarios. 
It can be seen that the KL divergence is sensitive to the value of $\rho$ of the proposed DAHC model, 
and the optimal values minimizing the KL divergence are $\rho^*=0.997,~0.997~\text{and}~0.996$ for Scenario 1 on 3.5GHz band, Scenario 1 on 60GHz band and Scenario 2 on 60GHz band respectively. 
Note that the KL divergence measures the distance between two distributions, the minimum KL divergence means the best match between the experiment and the SDP3 data simulator. 
Note that the samples in Fig.~\ref{fig:KLD}(b) and (c) are obtained with the same carrier frequency but different scenarios. 
This demonstrates the interference from non-target person: it leads to smaller value of calibrated hyper-parameter $\rho$, thus stronger level of sensing uncertainty. 
In all the three figures Fig.~\ref{fig:KLD}(a-c), it can be observed that the KL divergence between the experiment and the ray-tracing model is constant and significantly larger than the calibrated DAHC model. 
This demonstrates that it is necessary to include the sensing uncertainty in the channel model to capture the potential interference from non-target motions. 

In order to further justify the superior performance of the proposed DAHC model in sensing data generation, 
the grayscale distribution of spectrogram is compared in Fig.~\ref{fig:KLD}(d-f). 
Specifically, the spectrograms from experiment, ray-tracing model and DAHC model are generated for different carrier frequencies and scenarios. 
Then, the PMF of grayscale levels is calculated for each spectrogram. It can be observed that the grayscale PMFs from the experiment and proposed DHAC model match very well, 
and ray-tracing model fails to match the real experiment. This coincides with the comparison of KL divergence in Fig.~\ref{fig:KLD}(a-c). 

\subsection{Recognition Accuracy}

In this part, we continue to show that the sensing dataset generated via SDP3 data simulator with the DAHC model could achieve the recognition accuracy close to the experiment dataset. 
The hyper-parameter $\rho$ calibrated in the above part is adopted. The simulation dataset via ray-tracing model is also investigated as a comparison. 

Particularly, 600 samples of sensing signals are picked up from experiment, proposed SDP3 data simulator, ray-tracing-based simulator respectively. 
Their spectrograms are used to train the DSN in Section \ref{subsec:dsn} respectively. 
The training of DSN is implemented via Momentum optimizer with a learning rate of 0.06 and a mini-batch size of 300. 
The recognition accuracies of the trained DSNs are tested by other samples of the datasets. As a result, the recognition accuracy versus training epoch is illustrated in Fig.~\ref{fig:Comp}. 
It is observed that compared with the ray-tracing-based simulator, the curve of the SDP3 data simulator is closer to that of experiment. 
For example, it is shown that the accuracy gap are 2\% versus 5.33\% at 3.5GHz and 4\% versus 6.7\% at 60GHz respectively. 
This demonstrates that the DAHC model can simulate the motion recognition performance in real experiment better than the ray-tracing model. 
Hence, the dataset generated by the SDP3 data simulator could be used for the analysis of sensing-communication performance tradeoff. 
Thus, the significant effort of extensive real scenario experiment can be saved.

\subsection{Verification of SDP3 Performance Predictor}

The performance of the SDP3 performance predictor, as proposed in Section \ref{subsec:accuracy_model}, 
is demonstrated in this part, where the motion recognition accuracy versus the number of sensing slots $N_s$ is approximated analytically. 
We adopt the samples of received sensing signals generated by SDP3 data simulator. Both 3.5GHz and 60GHz bands are considered and five different motions, including 
child/adult standing, child walking, child pacing, adult walking and adult pacing, are classified. 
The samples in the dataset are generated with $N_s=1000$. 
The received signals of the first 200, 300, 400, 500, 600, 1000 sensing time slots are used for DSN training and accuracy testing, respectively, 
yielding the motion recognition accuracies for different values of $N_s$. 
Specifically, the training procedure is implemented via Adam optimizer with a learning rate of $0.01$ and a mini-batch size of $500$.
The convergence of the training loss for all the values of $N_s$ is shown in Fig.~\ref{fig:Training}(a).
Then the trained models are tested by $500$ new samples respectively, 
and the corresponding recognition accuracies can be obtained as shown in Fig.~\ref{fig:Training}(b). 
It can be observed that larger $N_s$ leads to better recognition accuracy. 

Hence, the curving fitting elaborated in Section \ref{sec:predictor} can be proceeded. 
The optimized tuning parameters of all the curve models are list in Table \ref{table_fitting}, 
and their comparison with the simulated accuracies is illustrated in Fig.~\ref{fig:Training}(c). 
It can be observed that the models of pow$_3$, log power and exp$_4$ match the simulated accuracy very well. 
This is because they possess higher curvature compared with other models. 
With the best-fitting model pow$_3$, the tuning parameters $\alpha$, $\beta$ and $\gamma$ are obtained by minimizing the MSE as in (\ref{fitting1}), yielding $(\alpha,\beta,\gamma)=(61906, 2.4297, 0.9499)$. 
Thus, the relation between the recognition accuracy and number of sensing slots is approximated as 
\begin{align}
	A = 0.9499-61906\times N_s^{-2.4297},  \label{approx_accuracy}
\end{align}

\begin{table}[!t]
	\centering
	\renewcommand\arraystretch{1.1}
	\caption{Parametric Learning Curve Models Fitting Results}
	\begin{center}
		\begin{tabular}{|c|c|c|}
		\hline
		\hline
		\textbf{Name} & \textbf{Parameters} & \textbf{MSE} \\\hline
		vapor & \multirow{2}{*}{\tabincell{c}{$\alpha=0.0117,\beta=-44.5180$}} & 
		\multirow{2}{*}{0.0017} \\
		pressure & & \\\hline
		pow$_3$   & \tabincell{c}{$\alpha=6.1906e4,\beta=2.4297$\\$\gamma=0.9499$} & 3.383e-4 \\\hline
		log power   & \tabincell{c}{$\alpha=0.9460,\beta=4.7438$\\$\gamma=-2.9235$} & 3.7018e-4 \\\hline
		exp$_4$  & \tabincell{c}{$\alpha=2.9129,\beta=6.9568$\\$\gamma=0.2082,\epsilon=0.9576$} & 5.4693e-4 \\\hline
		log log  & \multirow{2}{*}{\tabincell{c}{$\alpha=0.2347,\beta=1.0423$}} & 
		\multirow{2}{*}{0.0038} \\
		linear  & & \\\hline
		ilog2  & \tabincell{c}{$\alpha=3.5228,\beta=1.4863$} & 0.003 \\\hline
		pow$_4$  & \tabincell{c}{$\alpha=98.4911,\beta=-9.5892e3$\\$\gamma=1.2176,\epsilon=0.1117$} & 0.0065 \\\hline
		\hline
		\end{tabular}
	\end{center}
        \label{table_fitting}
\end{table}

\begin{figure}[!t]
	\centering
\includegraphics[width=85mm]{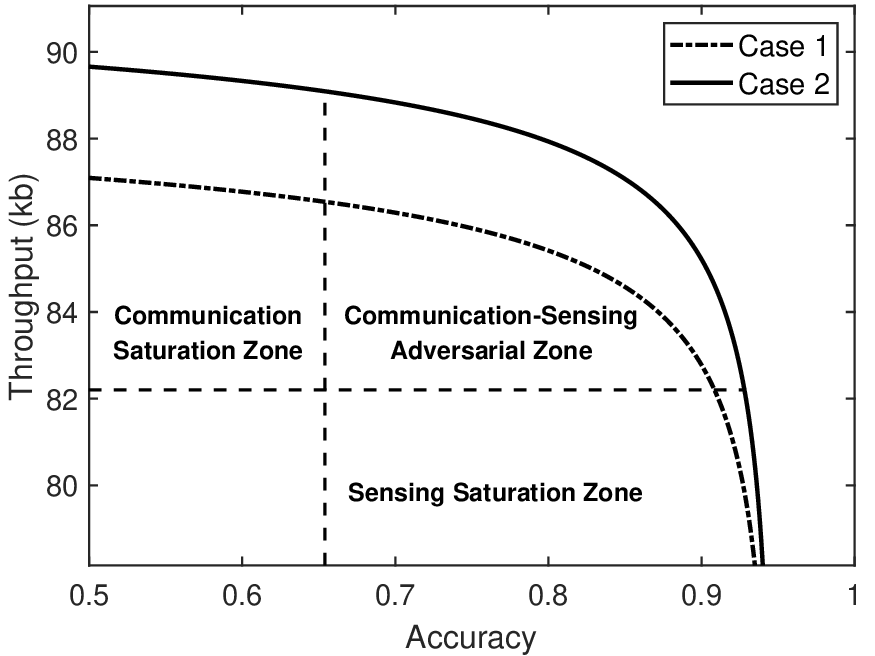}
		\caption{The A-T region for case 1 and 2 when $K=3$, $P=1$ Watt, $T_o=50\mathrm{\mu s}$, $T_s=250\mathrm{\mu s}$ and $N=4000$.}
		\label{Region} 
\end{figure}

\subsection{Sensing-Communication Tradeoff}

Finally, the SDP3 performance optimizer is utilized to illustrate the tradeoff between sensing and communication 
performance. The result with one sensing target and K=3 communication receivers is shown in Fig. \ref{Region}, 
where the carrier frequency is 3.5GHz. 
Two distributions of receivers are considered: 
the communication receivers are at the locations (1.5,3,1), (2.5,1,1) and (0.5,1,1) for the case 1, 
and at (1.5,2.8,1), (2.3,1,1) and (0.7,1,1) for the case 2. 

It can be observed that the accuracy-throughput (A-T) region of case 1 is a subset of that of case 2. 
This is because of the better communication channel in case 2. 
Both of two regions consist of three zones: 1) sensing saturation zone (bottom); 2) communication saturation zone (left); and 3) sensing-communication adversarial zone (upper right).
In the sensing saturation zone, reducing the communication performance can hardly improve the sensing performance, but a slight decrease of the sensing performance can significantly improve the communication throughput.
This is because excessive wireless resources have been allocated to sensing task.
The situation is the opposite in the communication saturation zone.
Therefore, for a practical ISAC system, it is not desirable to enter the sensing or communication saturation zone.
On the other hand, both sensing and communication performance varies sensitively with respect to each other in the sensing-communication adversarial zone. 
This is because the wireless resource allocation between the two functionalities is balanced. 
The proposed accuracy-throughput region analysis facilitates the search of the best Pareto point with heterogeneous quality-of-service requirements.

\section{Conclusion}

This paper proposes an SDP3 framework for the study of sensing-communication tradeoff with the particular application of human motion recognition. 
Specifically, the SDP3 data simulator with a data-driven hybrid channel model is proposed to generate the received sensing signals in a virtual environment. 
The SDP3 performance predictor is then introduced to approximate the motion recognition accuracy via analytical expression with the simulated dataset of sensing signals. 
Finally, the recognition accuracy and communication throughput tradeoff is characterized by the SDP3 performance optimizer. 
It is demonstrated that the dataset generated by the SDP3 data simulator matches the experiment dataset in KL divergence, 
grayscale PMF and motion recognition accuracy. Hence, the sensing-communication tradeoff can be investigated without extensive experiments. 
It is also shown that the sensing and communication performance is balanced in the sensing-communication adversarial zone of the A-T region, where both performance varies sensitively with respect to each other. 

\appendices

\section{Proof of Theorem 1} \label{appx:Theorem1}
Appling the water-filling policy to the power constraints, the optimal worst communication throughput 
among all receivers is 
\begin{align}
R = \mathop{\mathrm{min}}_{k=1,\cdots,K}
\frac{N_{c,k}T_s}{T_o}\sum\limits_{m=1 \atop \gamma_{k,m}\geq\gamma_{k,0}}^{N} \log_{2} \left(\frac{\gamma_{k,m}}{\gamma_{k,0}}\right), \label{10}
\end{align}
where
\begin{align}
\sum\limits_{m=1}^{M}\left(\frac{1}{\gamma_{k,0}}-\frac{1}{\gamma_{k,m}}\right) = 1, \label{water}
\end{align}
Then the $\mathcal{P}_1$ could be
\begin{align}
\mathcal{P}_1:
\mathop{\mathrm{max}}_{\substack{A,R,N_s,\mathbf{N_c}}}
~~&(A,R),
\nonumber\\
\mathrm{s.t.}~~~~
&\eqref{10} \nonumber\\
&A=\Theta(C),~N_s+\sum\limits_{k=1}^{K}N_{c,k}=N.
\label{Problem_2}
\end{align}
Given a fixed $N_s=N_s^*$, the optimal $\mathbf{t}$ is derived using KKT optimality conditions.
Specifically, the Lagrangian of problem $\mathcal{P}$ is given by
\begin{align}
L&=-R^*+\sum_{k=1}^K\eta_k\left(R^*-
\frac{N_{c,k}^*T_s}{T_o}\sum\limits_{m=1 \atop \gamma_{k,m}\geq\gamma_{k,0}}^{N} \log_{2} \left(\frac{\gamma_{k,m}}{\gamma_{k,0}}\right)
	\right) \nonumber\\
&\quad+\varrho\left(N_s^*+\sum\limits_{k=1}^{K}N_{c,k}^*-N\right),
\label{Problem1}
\end{align}
where $\{\eta_k\geq 0, \varrho\}$ are Lagrangian multipliers.
By letting $\partial L/\partial R^*=0$ and $\partial L/\partial N_{c,k}^*=0$, we have
\begin{align}
&\sum_{k=1}^K\eta_k=1,\quad -\frac{\eta_kT_s}{T_o} \sum\limits_{m=1 \atop \gamma_{k,m}\geq \gamma_{k,0}}^{N} \log_{2} \left(\frac{\gamma_{k,m}}{\gamma_{k,0}}\right)+\varrho=0. \label{kkt}
\end{align}
Now we will prove that $\eta_k\neq 0$ for any $k$ by contradiction.
In particular, assume that $\eta_j=0$ for some $j$.
Putting $\eta_j=0$ into the second equation of \eqref{kkt} yields $\varrho=0$.
Putting $\varrho=0$ into the second equation of \eqref{kkt} with $k\neq j$ yields $\eta_k=0$ for any $k\neq j$.
Lastly, based on $\eta_j=0$ and $\eta_k=0$ for $k\neq j$, we have $\sum_{k=1}^K\eta_k=0$. This contradicts to $\sum_{k=1}^K\eta_k=1$ of the first equation of \eqref{kkt}.
Therefore, $\eta_k\neq 0$ for any $k$.
Using the above result and the complementary slackness condition\\
\begin{align}
\eta_k\left(R^*-\frac{N_{c,k}^{*}T_s}{T_o}\sum\limits_{m=1 \atop \gamma_{k,m}\geq\gamma_{k,0}}^{N} \log_{2} \left(\frac{\gamma_{k,m}}{\gamma_{k,0}}\right)
\right)=0,
\end{align}
the following equality is obtained
\begin{align}
N_{c,1}^* \sum\limits_{m=1 \atop \gamma_{1,m}\geq \gamma_{1,0}}^{N} \log_{2} \left(\frac{\gamma_{1,m}}{\gamma_{1,0}}\right)=\cdots
\nonumber\\=N_{c,K}^* \sum\limits_{m=1 \atop \gamma_{K,m}\geq \gamma_{K,0}}^{N} \log_{2} \left(\frac{\gamma_{K,m}}{\gamma_{K,0}}\right). \label{Eq}
\end{align}
Combining the above result with the constraint $N_s^*+\sum\limits_{k=1}^{K}N_{c,k}^*=N$ yields\\
\begin{align}
&N_{c,k}^*=
\frac{N-N_s^*}{\Bigg(\mathop{\sum}\limits_{k=1}^K\frac{1}{\sum\limits_{m=1 \atop \gamma_{k,m}\geq \gamma_{k,0}}^{N} \log_{2} \left(\frac{\gamma_{k,m}}{\gamma_{k,0}}\right)}\Bigg)\sum\limits_{m=1 \atop \gamma_{k,m}\geq \gamma_{k,0}}^{N} \log_{2} \left(\frac{\gamma_{k,m}}{\gamma_{k,0}}\right)}.
\end{align}
According to \eqref{10}
\begin{align}
A^*& = \Theta(N_s^*), \label{A*}
\end{align}
\begin{align}
R^*& = \frac{(N-N_s^*)T_s}{\sum\limits_{k=1}^K T_o \left(\sum\limits_{m=1 \atop \gamma_{k,m}\geq \gamma_{k,0}}^{N} \log_{2} \left(\frac{\gamma_{k,m}}{\gamma_{k,0}}\right)\right)^{-1}}. \label{R*}
\end{align}
Rearranging equations \eqref{A*}--\eqref{R*}, the proof is completed.

\end{document}